\documentclass[journal]{IEEEtran}

\usepackage{amsmath}
\usepackage{amsthm}
\usepackage{array}
\usepackage{fixltx2e}
\usepackage{stfloats}
\usepackage{url}
\usepackage[table]{xcolor}
\usepackage{cite}
\usepackage{amsmath,amssymb,amsfonts}
\usepackage{amsbsy}
\usepackage{algorithm}
\usepackage{graphicx}
\usepackage{graphics}
\usepackage{textcomp}
\usepackage{xcolor}
\usepackage{mathtools,amssymb,lipsum,nccmath,bm,amsmath}
\usepackage{cuted}
\usepackage{array}
\usepackage{ragged2e}
\usepackage[short]{optidef}
\usepackage{amsmath,algorithm,tabularx}
\usepackage{algpseudocode}
\usepackage{float}
\usepackage{tikz}
\usepackage{subcaption}

\newtheorem{remark}{Remark}

\usepackage{mathtools}

\usepackage{tikz}
\usetikzlibrary{shapes, arrows.meta, positioning}

\IEEEoverridecommandlockouts
\allowdisplaybreaks[2]

\begin{document}
\title{Real-Time Power System Event Detection:  \\
A Novel Instance Selection Approach}

\author{
    \IEEEauthorblockN{Gabriel Intriago,~\IEEEmembership{Student Member,~IEEE} and Yu Zhang,~\IEEEmembership{Member,~IEEE}}

    \thanks{G. Intriago is with the Department of Electrical and Computer Engineering at the University of California Santa Cruz, USA (e-mail: gintriag@ucsc.edu). Also, he is with the Math Department at the Escuela Politécnica Nacional, Ecuador (e-mail: gabriel.intriago01@epn.edu.ec).}

    \thanks{Y. Zhang is with the Department of Electrical and Computer Engineering at the University of California, Santa Cruz, USA (e-mail: zhangy@ucsc.edu).}

    }

\maketitle

\begin{abstract}
Instance selection is a vital technique for energy big data analytics. It is challenging to process a massive amount of streaming data generated at high speed rates by intelligent monitoring devices. Instance selection aims at removing noisy and bad data that can compromise the performance of data-driven learners. In this context, this paper proposes a novel similarity based instance selection (SIS) method for real-time phasor measurement unit data. In addition, we develop a variant of the Hoeffding-Tree learner enhanced with the SIS for classifying disturbances and cyber-attacks. We validate the merits of the proposed learner by exploring its performance under four scenarios that affect either the system physics or the monitoring architecture. Our experiments are simulated by using the datasets of industrial control system cyber-attacks. Finally, we conduct an implementation analysis which shows the deployment feasibility and high-performance potential of the proposed learner, as a part of real-time monitoring applications.
\end{abstract}


%--------------------INTRODUCTION------------------------------
%------------------------------------------------------------------
\section{Introduction}
\IEEEPARstart{T}{he} undergoing deployment of intelligent monitoring devices has led to significant transformations of electric power systems from purely physical systems into cyber-physical systems \cite{AkhavanHejazi2018}. Breakthroughs in the technology of storage, sensors and communications have enabled devices to operate in a more automated fashion with very limited human interventions.  This leads to fast and continuous energy data streams \cite{Aggarwal2007}. The need of increasing situational awareness of power grids and the ever-growing large volumes of data streams pose a series of new challenges for traditional data analysis methods. In this context, it is indispensable to design an effective instance selection policy that chooses the most relevant subset from the original stream. Bad or no instance selection may incur performance degradation of machine learning algorithms, let alone the prohibited processing times for real-time environments.

Furthermore, energy data distributions are often time varying that is also known as concept drift \cite{Gama2014}. A change in data distribution which may lead to sub-optimal decisions and inaccurate predictions. An intelligent data instance selection approach must be robust against potential concept drift, which can occur for a few unexpected reasons. For example, an intrusion attack to the control room, a line to ground fault on a specific bus, or communication interruption of a phasor measurement unit (PMU).

\subsection{Motivation}
\begin{figure}
    \centering
    \includegraphics[width=0.9\linewidth]{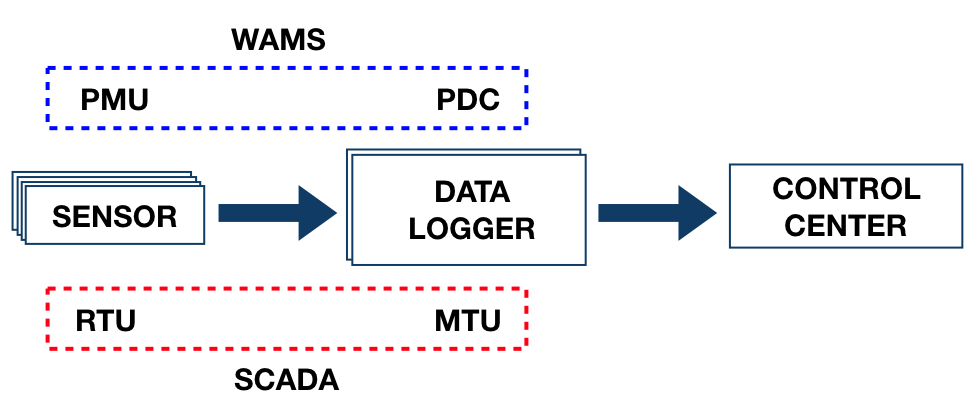}
    \caption{A generic diagram of an infrastructure for power systems monitoring. The wide area monitoring systems (WAMS) architecture relies on phasor measurement units (PMUs) and phasor data concentrators (PDCs) to generate and acquire real-time data respectively. The supervisory control, and data acquisition (SCADA) systems obtain data from remote terminal units (RTUs) which are collected and sent to the control center by master terminal units (MTUs).}
    \label{monitoring}
\end{figure}

Fig. \ref{monitoring} shows a generic diagram of a power system monitoring architecture. As we can see, an array of sensors first send measurements to a logger that collects and organizes the data. Then, the loggers send the organized data to a control center, which help a system operator to take preventive and corrective actions. SCADA and WAMS are the two widely used technologies for the real-time system monitoring. The SCADA sensors are the RTUs, the loggers are the MTUs, and the measurements are generated typically in the order of seconds. For the WAMS technology, the sensors are the PMUs, the loggers are the PDCs, and the measurements are generated in the order of milliseconds. Note that a PMU can generate about one Gigabyte data per day. An RTU can create a tabular dataset with more than 2 million rows per month. 
The situation is further aggravated when the monitoring architecture have several data loggers and a large number of sensors. Then, it becomes challenging to store and process this data in a single pass. In this context, it is essential to model data as a stream rather than as a static batch and find a way to select a reduced-size optimal subset from the original data.

\subsection{Related Literature}
A prototype-based instance selection is utilized for data reduction in classification tasks \cite{Ougiaroglou20142}. When data evolve, it is hard to determine a fixed number of instance neighbors \cite{Domeniconi2002}. A unifying windowing and instance selection framework under concept drift are explored \cite{Zliobaite2011}. The authors propose an approach that generalizes training set selection by combining space and time similarity. In \cite{Tsymbal2008}, the authors propose a dynamic ensemble method that evaluates the base learners using cross-validation on a set of the target instance's nearest neighbors. An instance selection method was originally designed to use with a support vector machine (SVM) as the base learner \cite{10.5555/645529.657791}. The goal is to choose a window size that reduces the generalization error on testing examples. However, such a design is not applicable for a real-time scenario where data are modeled as a continuous stream.

It is possible to group sensors with high correlation to maintain real-time clusters integrated with a predictive model that adapts with the most recent information \cite{Gama2007StreamBasedEL}. More recently, the concept of a common path is applied for events classification in smart grids \cite{Pan2015}. In \cite{Dahal2015}, the authors use a Hoeffding adaptive tree (HAT) to classify between normal operation and electrical faults. A combination of non-nested generalized exemplars (NNGE) with the state extraction method (STEM) is proved to be an effective classifier of events and cyber-attacks in real-time \cite{Adhikari2018}. In \cite{Mrabet2019}, the authors propose a transfer learning HAT model with the adaptive windowing (ADWIN) drift detector. Their approach transfers knowledge between four datasets, where each dataset is corresponding to a specific frequency oscillation of the power system. HAT with two change detectors (drift detection method (DDM) and ADWIN) is used in an event and intrusion classification environment \cite{Adhikari20182}. The combination of HAT+DDM+ADWIN can be enhanced by unsupervised learning to improve the performance for power system event classification \cite{Intriago2021}.

\subsection{Contribution}
The main contributions of this paper are listed as follows:
\begin{itemize}
\item We propose an effective streaming instance selection (SIS) method for events and intrusion classification in power systems. Our focus is to select the most similar instances to the target instance by using a spatio-temporal distance function that adapts to the scaling of the PMU measurements. In general, the proposed SIS method can find the optimal subset of data instances, even in a non-stationary streaming environment.

\item We develop the Hoeffding Adaptive Tree (HAT) combined with the SIS method, namely HAT+SIS. The proposed classifier proves to work under the streaming learning requirements while keeping a low memory consumption and running time. 

\item Extensive simulation results corroborate the merits of the proposed classifier, which is used to test for 37 power system events (e.g., disturbances and cyber-attacks). We investigate scenarios that impact the system physics and its monitoring infrastructure, such as similar events with different loading schemes, the sudden disconnection of a PMU, and events with similar measurements.
\end{itemize}

\section{Preliminaries}
\subsection{Concept Drift and Feature Drift}
In streaming data learning lexicon, concepts are defined as the target information that a model aims to predict \cite{Gomes2019}. Data streams are inherently infinite, temporal and dynamic. The data distribution may evolve over time while the mapping between instances and targets can be time-varying. Such a situation gives rise to the phenomenon of \textit{concept drift} \cite{Gama2014}. 

As a particular case of concept drift, feature drift occurs when a subset of features becomes irrelevant to the learning task \cite{Gomes2019}. In this work, the features are the PMU measurements such as voltages, currents, impedances, etc. Feature drift in the context of power systems includes e.g., (i) the removal of an existing PMU from the monitoring system; (ii) less informative voltage magnitude measurements due to bad data injection; and (iii) changes in the PMU measurements due to cyber-attack events.

\subsection{Classification for Data Streams}
Classification for data streams inherits large numbers of  problems in batch learning. There are also new challenges such as one-pass learning, limited processing time/memory, and changes in data distribution.

In this work, we focus on data stream classification. Let $\left\{(\bm{x}_t,y_t) \right\}_{t=1}^\infty$ denote a data stream that contains a set of labelled instances. 
At time $t$, $\bm{x}_t \in \mathbb{R}^m$ denotes the vector of $m$ features while $y_t$ is the corresponding class label. Let $\mathcal{X}$ represent the entire feature space and $\mathcal{Y}$ the class space. A classification algorithm learns a mapping $f: \mathcal{X} \mapsto \mathcal{Y}$ such that it can be used to  predict the class label for a new instance. Batch classification can afford to load all the data into memory. In contrast, stream classification is a one-pass strategy, meaning that the processed instances are automatically discarded or stored temporarily.

A learning model should meet the following criteria to be compliant with data stream learning \cite{Bifet_datastream}:
\begin{itemize}
\item Learn an instance at a time, and inspect it at most once.
\item The model must use a limited amount of memory.
\item The working time is limited.
\item The model must be able to predict at any time.
\end{itemize}
% Fig. \ref{learningcycle} shows the framework of the instance selection combined with streaming learning.

%--------------------METHODOLOGY------------------------------
%------------------------------------------------------------------
\section{Methodology}
In this section, we first present the concept of spatio-temporal similarity. Next, we discuss the streaming learning setup and the details of our proposed streaming instance selection. Finally, we briefly describe the base learners.

\subsection{Linear Spatio-temporal Similarity}
\begin{figure}[H]
\centering
\includegraphics[width=0.6\linewidth]{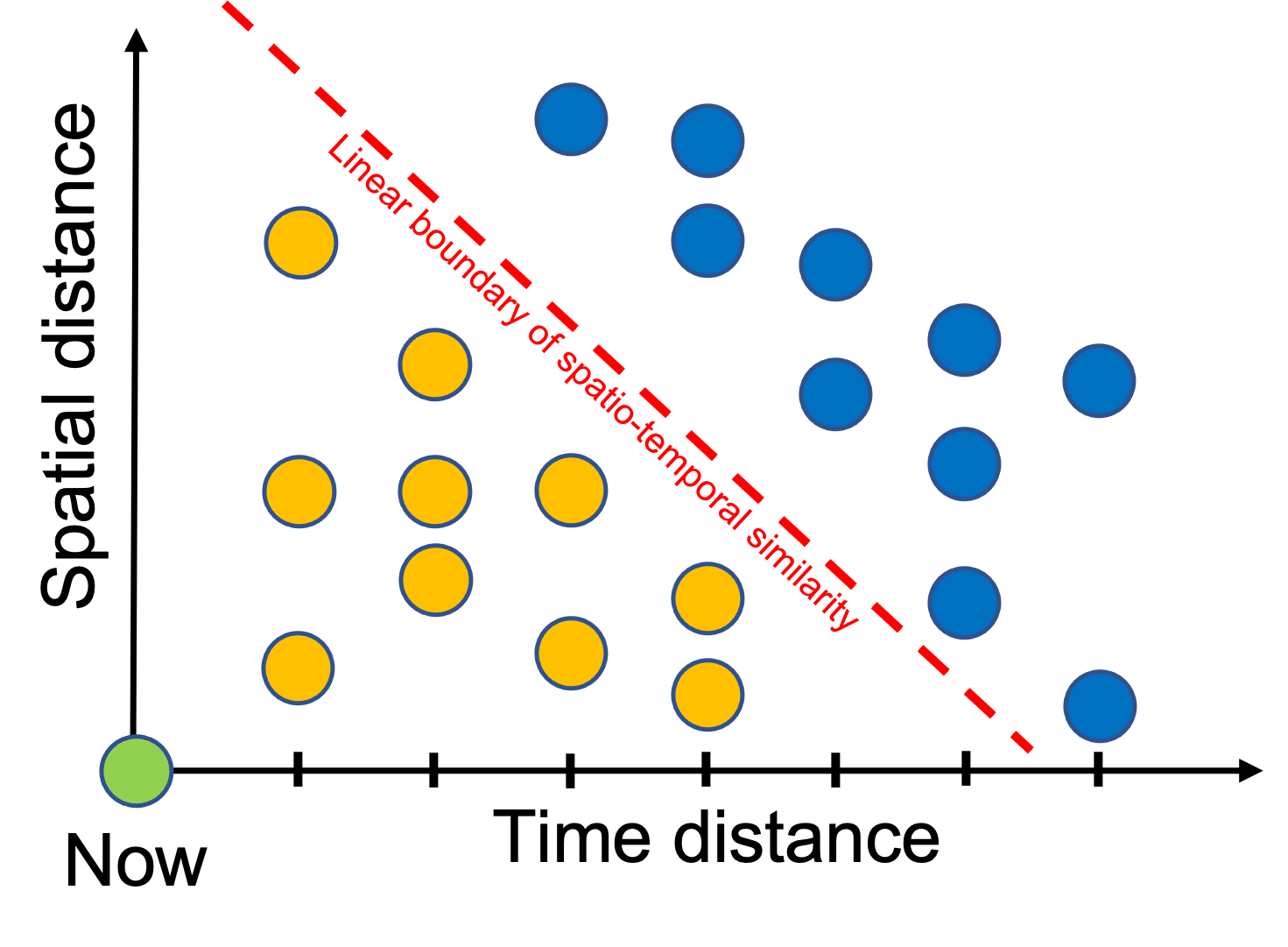}
\caption{Linear combination of time and spatial distances. Each circle corresponds to a historical instance. The green circles are the target instance while the yellow ones are the similar instances to the targets because they have the least spatio-temporal distance determined by the red similarity boundary.}
\label{spatiotemporal}
\end{figure}

Our proposed instance selection technique for data streams is based on similarity among instances. The similarity captures how comparable are a pair of instances. A way to measure similarity is by using a distance function. The concept of similarity is inversely related to the idea of distance \cite{Kordos2019}. In other words, the smaller the distance between instances, the more similar they are. Usually, the term \textit{distance} is associated with distance in space. However, distance can not only be defined in space but in time. Moreover, distance can be defined as a function of both time and space. We refer to the term distance as the spatio-temporal distance between two instances as suggested in \cite{Zliobaite2011} and shown in Fig. \ref{spatiotemporal}.

\begin{remark}
Potentially accurate results can be obtained using a spatio-temporal distance for power systems events detection. For example, an instance with a low time distance and a high spatial distance may represent an abrupt change in the concept, such as transitioning from normal operation to double line fault. An instance with a low spatial distance and high time distance may indicate the evolution of a concept such as a single line fault under two different loading conditions.
\end{remark}

%\begin{remark}
%Potentially accurate results can be obtained using a spatio-temporal distance for power systems events detection. For example, an instance with a low time distance and a high spatial distance may represent an abrupt change in the concept, such as transitioning from normal operation to double line fault. An instance with a low spatial distance and high time distance may indicate two situations: 1) the presence of a reoccurring concept such as the seasonal variation in load demand, or 2) the evolution of a concept such as a single line fault under two different loading conditions.
%\end{remark}

Let $\bm{x}_t \in \mathbb{R}^m$ be the target instance whose class is to be predicted, $\bm{x}_i \in \mathbb{R}^m$ an already observed instance, $T(\cdot)$ a distance function in time, and $S(\cdot)$ a distance function in space. More formally, the spatio-temporal distance between $\bm{x}_t$ and $\bm{x}_i$ is defined as the following linear relation:
\begin{align}
D(\bm{x}_t,\bm{x}_i) &= T(\bm{x}_t,\bm{x}_i) + S(\bm{x}_t,\bm{x}_i)
\label{combdist}
\end{align}
To simplify the notation, we refer to the distance from $\bm{x}_i$ to the target instance $\bm{x}_t$ as $D_{t-i}$. In Figure \ref{spatiotemporal}, we illustrate the concept of linearly combining the spatial and time distances. The are multiple candidate functions for $S(\cdot)$, for example, the Euclidean distance, Manhattan distance, cosine similarity distance, or any other existing function that measures spatial distance. In this work, we choose the Euclidean distance:
\begin{align}
S(\bm{x}_t,\bm{x}_i) = \left\|\bm{x}_t - \bm{x}_i\right\|_2
\label{spatialdist}
\end{align}
Assuming uniformly spaced time intervals, and considering the $N$ most recent observed instances, we choose the distance in time defined as a linear function of the time indices:
\begin{align}
T(\bm{x}_t,\bm{x}_i) = \frac{\left|t - i \right|}{N}
\label{timedist}
\end{align}
Other more complex time distances can be chosen, for example the exponential function $T(\bm{x}_t,\bm{x}_i) = e^{\left|t - i \right|}$. Such choice gives more importance to recent instances, however we leave this and other complex spatial and time distance functions for future work.

A weight can be assigned to the Euclidean distance function using a fixed or cross-validation strategy \cite{Zliobaite2011}. This work gives a time-variant weight $\alpha_t$ to the Euclidean distance function to reduce the impact of an inappropriate range of features values. To do so, we use a scaling function $s: \mathbb{R}^m \mapsto \mathbb{R}^m$ that transforms the feature domain of all instances in such a way that the values of the features are on a similar scale. Specifically, we scale the instances so that the values of the features have zero mean and unit variance. At each time step, a running mean and running variance are maintained. The scaling is different from the offline scaling  because the exact means and variances are unknown beforehand \cite{StandardScaling2021}. In this way, the Euclidean distance becomes:
\begin{align}
S(\bm{x}_t,\bm{x}_i) = \left\|s(\bm{x}_t) - s(\bm{x}_i)\right\|_2
 = \alpha_t\left\|\bm{x}_t - \bm{x}_i\right\|_2
\label{spatialdist}
\end{align}

\begin{remark}
Power systems measurements make the spatio-temporal distance vulnerable to improper features scaling. For example, data packets from PMUs come from different units. Therefore, measurements exhibit different orders of magnitude.
\end{remark}

\subsection{Streaming Learning Framework}
\begin{figure}[H]
    \centering
    \includegraphics[width=0.8\linewidth]{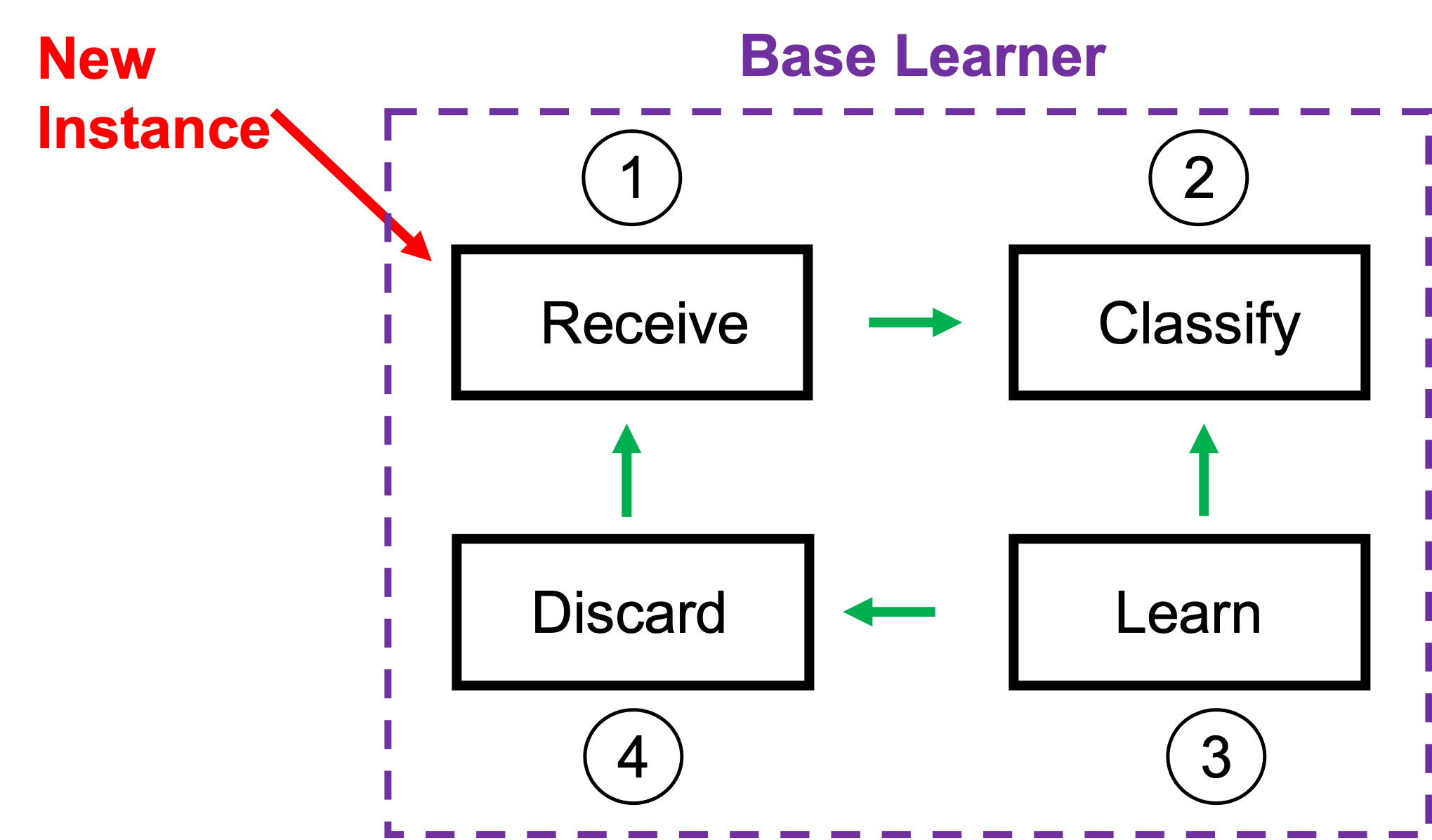}
    \caption{Streaming learning framework with instance selection.}
    \label{learningcycle}
\end{figure}

To be compliant with streaming learning framework, we follow the rules of \textit{Prequential evaluation} \cite{Gama2012}. Essentially, prequential evaluation has two major stages: test and train. In the test stage, the base learner predicts the class of the next available instance from the stream. Right after testing, the model metrics are updated. The base learner processes the instance in the training stage and then updates its structure and statistics. In our work, the training stage is governed by our proposed streaming instance selection methods, which we will explain in detail in the next section. Algorithm \ref{streamlearning} shows the streaming learning process under the rules of prequential evaluation. Fig. \ref{learningcycle} depicts the streaming learning framework with instance selection. 

\begin{algorithm}[t]
\caption{Streaming Learning}
\label{streamlearning}
\begin{algorithmic}[1]
\Require $\mathcal{M} = \left\{(\bm{x}_t,y_t)\right\}_{t=1}^\infty$, base learner $\mathcal{L}$
\For{$t=1,2,3,\dots$}
\State $\triangleright$ Scale the feature vector $\bm{x}_t$ incrementally
\State $\triangleright$ Test $\mathcal{L}$ with $\left(\bm{x}_t,y_t \right)$
\State $\triangleright$ Update the model metrics
\State $\triangleright$ Train $\mathcal{L}$ with SIS using Algorithm \ref{SIS-C}
\EndFor
\end{algorithmic}
\end{algorithm}

\begin{algorithm}[t]
\caption{Streaming Instance Selection (SIS)}
\label{SIS-C}
\begin{algorithmic}[1]
\Require base learner $\mathcal{L}$, set of most recent instances $\mathcal{R} = \{(\bm{x}_{t-i},y_{t-i})\}_{i=1}^{N}$
% \ENSURE $y = x^n$
\State $\triangleright$ Resize the dimension of instances from set $\mathcal{R}$ according to the dimension of $\bm{x}_t$
\State $\triangleright$ Reorder instances from set $\mathcal{R}$ using Algorithm \ref{reordering}
\State $\triangleright$ Reset the base learner $\mathcal{L}$
\State $\triangleright$ Search the optimal size for sliding window, and train the base learner $\mathcal{L}$ using Algorithm \ref{setsearch}
\State \textbf{Return} $\mathcal{L}$
\end{algorithmic}
\end{algorithm}

\begin{algorithm}[t]
\caption{Instance Reordering}
\label{reordering}
\begin{algorithmic}[1]
\Require Target instance $\bm{x}_t$, set of most recent instances $\mathcal{R} = \{(\bm{x}_{t-i},y_{t-i})\}_{i=1}^{N}$
\For{$i=1,\dots,N$}
\State $\triangleright$ Compute the distance $D_{t-i}$ according to \eqref{combdist}
\EndFor
\State $\triangleright$ Sort the distances in ascending order $D_{g(1)} < D_{g(2)} < \dots < D_{g(N)}$
\State $\triangleright$ Build the natural-valued function $g:\mathcal{U}\mapsto \mathcal{V}$, where $\mathcal{U} = \{1,2,\dots,N\}$, and $\mathcal{V} = \{t-1,t-2,\dots,t-N\}$
\State \textbf{Return} $g$
\end{algorithmic}
\end{algorithm}

\subsection{The SIS Method}
This section presents the proposed streaming instance selection (SIS) method. Let $\bm{x}_t$ be the target instance, $\mathcal{L}$ be the base learner, and $\mathcal{R} = \{(\bm{x}_{t-i},y_{t-i})\}_{i=1}^{N}$ the set containing the most recent observed instances. Let $g:\mathcal{U}\mapsto \mathcal{V}$ denote the one-to-one natural-valued function, where $\mathcal{U} = \{1,2,\dots,N\}$, and $\mathcal{V} = \{t-1,t-2,\dots,t-N\}$. SIS begins by sorting the set of observed instances according to their distance to the target instance $\bm{x}_t$. The base learner is reset to forget what was learned at the previous time step. The reset allows the model to adapt to concept drifts. Note that the effectiveness of bypassing concept drifts depends on $N$. Then, the base learner is trained incrementally with a new set of indices $\{g(1), g(2),\dots,g(N)\}$ obtained from the sorting procedure, driving the base learner to learn first with the most similar instances to the target. The size of the window $\mathcal{W}$ is chosen dynamically, which is upper bounded by $ N$. SIS evaluates the trained base learner with a trial set containing the $k$ most recent instances. The trial set is indexed by time $\{t-1,t-2,\dots,t-N\}$, and not by the sorted indices $\{g(1),g(2),\dots,g(N)\}$. The training window $\mathcal{W}$ is found by using a warm restart to alleviate processing all the most recent instances in $\mathcal{R}$. SIS performs a local search around the previous best window size to find the new best size. Let $b$ be the previous best window size and $r$ a natural number. SIS searches the next best window's size in the interval $[l,u] \subseteq [1,N]$, where $l=b-r$ and $u=b+r$. Finally, SIS stops the search when the learner error on the trial set is less than a threshold $\epsilon$. The computational complexity of SIS is $\frac{1}{2}N(N-1) + N + 2rk = \mathcal{O}(rk + N^2)$. Note that it is desirable if $r \ll N/2$ and $k \ll N$ hold true.

\begin{algorithm}[t]
\caption{Optimal Window Search}
\label{setsearch}
\begin{algorithmic}[1] % $|\mathcal{R}| = N$
\Require $|\mathcal{R}| = N$, base learner $\mathcal{L}$, natural-valued function $g$, number of testing instances $k$, previous best window size $b$, natural number $r$, error threshold $\epsilon$
\State $\triangleright$ Set the upper and lower limits $u = b + r$, $l = b - r$
    \For{$i = 1:N$}
        \If{$i > u$}
            \State $\triangleright$ \textbf{break}
        \EndIf
        \State $\triangleright$ Train the base learner $\mathcal{L}$ with $(\bm{x}_{g(i)},y_{g(i)})$
        \If{$i < l$}
            \State $\triangleright$ \textbf{continue}
        \EndIf 
        \For{$j=1 \: : \: k$}
            \State $\triangleright$ Test base learner $\mathcal{L}$ with $(\bm{x}_j,y_j)$
            \State $\triangleright$ Update metric $\mathrm{learnerError}$
        \EndFor
        \If{$\mathrm{learnerError} < \epsilon$}
            \State $\triangleright$ $b = i$ and \textbf{break}
        \EndIf
    \EndFor
\State \textbf{Return} $\mathcal{L}$, $b$
\end{algorithmic}
\end{algorithm}

At each time step, SIS assigns a new index to the instances in $\mathcal{R}$ by solving the following optimization problem:
\begin{mini}
  {g:\mathcal{U}\mapsto \mathcal{V}}{\sum_{i=2}^N |D_{g(i)} - D_{g(i-1)}|}{}{}
  \addConstraint{\mathcal{U} = \{1,2,\dots,N\}}
  \addConstraint{\mathcal{V} = \{t-1,t-2,\dots,t-N\}}
  \label{reorderingopt}
\end{mini}

\begin{figure}
\centering
\begin{tikzpicture}[font=\scriptsize,node distance=0.5cm,thick]
% Start block
\node[draw,
    rounded rectangle,
    minimum width=1.5cm,
    minimum height=0.5cm] (block1) {START};
 
% Blocks
\node[draw,
    below=of block1,
    minimum width=3.5cm,
    minimum height=0.5cm
] (block2) {Scale the feature vector $\bm{x}_t$ incrementally};

\node[draw,
    below=of block2,
    minimum width=3.5cm,
    minimum height=0.5cm
] (block3) {Test $\mathcal{L}$ with $\left(\bm{x}_t,y_t \right)$};

\node[draw,
    below=of block3,
    minimum width=3.5cm,
    minimum height=0.5cm
] (block4) {Update the model metrics};

\node[draw,
    align=center,
    below=of block4,
    minimum width=3.5cm,
    minimum height=0.5cm
] (block5) {Resize the dimension of instances from set $\mathcal{R}$ according \\ to the dimension of $\bm{x}_t$};

\node[draw,
    align=center,
    below=of block5,
    minimum width=3.5cm,
    minimum height=0.5cm
] (block6) {Compute the distance $D_{t-i}$ according to \eqref{combdist}};

% Condition test 1
\node[draw,
    diamond,
    below=of block6,
    minimum width=1.5cm,
    inner sep=0] (block7) {i $<$ N};

\node[draw,
    align=center,
    below=of block7,
    minimum width=3.5cm,
    minimum height=0.5cm
] (block8) {Sort the distances in ascending order $D_{g(1)} < D_{g(2)} < \dots < D_{g(N)}$};

\node[draw,
    align=center,
    below=of block8,
    minimum width=3.5cm,
    minimum height=0.5cm
] (block9) {Build the natural-valued function $g:\mathcal{U}\mapsto \mathcal{V}$, where $\mathcal{U} = \{1,2,\dots,N\}$, \\ and $\mathcal{V} = \{t-1,t-2,\dots,t-N\}$};

\node[draw,
    align=center,
    below=of block9,
    minimum width=3.5cm,
    minimum height=0.5cm
] (block10) {Reset the base learner $\mathcal{L}$};

\node[draw,
    align=center,
    below=of block10,
    minimum width=3.5cm,
    minimum height=0.5cm
] (block11) {Set the upper and lower limits $u = b + r$, $l = b - r$};

% Condition test 2
\node[draw,
    diamond,
    below=of block11,
    minimum width=1.5cm,
    inner sep=0] (block12) {i $>$ u};

\node[draw,
    align=center,
    below=of block12,
    minimum width=3.5cm,
    minimum height=0.5cm
] (block13) {Train the base learner $\mathcal{L}$ with $(\bm{x}_{g(i)},y_{g(i)})$};

% Condition test 3
\node[draw,
    diamond,
    below=of block13,
    minimum width=1.5cm,
    inner sep=0] (block14) {i $<$ l};

\node[draw,
    align=center,
    below=of block14,
    minimum width=3.5cm,
    minimum height=0.5cm
] (block15) {Test base learner $\mathcal{L}$ with $(\bm{x}_j,y_j)$};

\node[draw,
    align=center,
    below=of block15,
    minimum width=3.5cm,
    minimum height=0.5cm
] (block16) {Update metric $\mathrm{learnerError}$};

% Condition test 4
\node[draw,
    diamond,
    below=of block16,
    minimum width=1.5cm,
    inner sep=0] (block17) {j $<$ k};

% Condition test 5
\node[draw,
    diamond,
    below=of block17,
    aspect=2,
    inner sep=0] (block18) {$\mathrm{learnerError}$ $<$ $\epsilon$};

\node[draw,
    align=center,
    left=of block18,
    minimum width=0.5cm,
    minimum height=0.5cm
] (block19) {$b=i$};

% Condition test 6
\node[draw,
    diamond,
    below=of block18,
    minimum width=1.5cm,
    inner sep=0] (block20) {i $<$ N};

% Condition test 6
\node[draw,
    diamond,
    below=of block20,
    minimum width=1.5cm,
    inner sep=0] (block21) {t $< \infty$ };

% Start block
\node[draw,
    rounded rectangle,
    below=of block21,
    minimum width=1.5cm,
    minimum height=0.5cm] (block22) {END};

%% Arrows
\draw[-latex] (block1) edge (block2)
    (block2) edge (block3)
    (block3) edge (block4)
    (block4) edge (block5)
    (block5) edge (block6)
    (block6) edge (block7)
    % (block7) edge (block8)    
    (block8) edge (block9)  
    (block9) edge (block10)
    (block10) edge (block11)
    (block11) edge (block12)
    (block12) edge (block13)
    (block13) edge (block14)
    (block14) edge (block15)
    (block15) edge (block16) 
    (block16) edge (block17)
    (block17) edge (block18)
    (block18) edge (block19)
    (block18) edge (block20)
    (block20) edge (block21);

\draw[-latex] 
    (block7) edge node[pos=0.4,fill=white,inner sep=2pt]{No}(block8)
    % [->] (block7.west) -| ++(-2.0,0) |- (block6.west);
    (block7) -| ++(-3.0,0) node[pos=0.1,fill=white,inner sep=0]{Yes} |- (block6);

\draw[-latex] 
    (block14) edge node[pos=0.4,fill=white,inner sep=2pt]{No}(block15)
    % [->] (block7.west) -| ++(-2.0,0) |- (block6.west);
    (block14) -| ++(3.0,0.0) node[pos=0.1,fill=white,inner sep=0]{Yes} |- (block12);

\draw[-latex] 
    (block17) edge node[pos=0.4,fill=white,inner sep=2pt]{No}(block18)
    % [->] (block7.west) -| ++(-2.0,0) |- (block6.west);
    (block17) -| ++(-2.5,0.0) node[pos=0.1,fill=white,inner sep=0]{Yes} |- (block15);

\draw[-latex] 
    (block18) edge node[pos=0.4,fill=white,inner sep=0pt]{Yes}(block19)
    (block18) edge node[pos=0.4,fill=white,inner sep=2pt]{No}(block20);
    % [->] (block7.west) -| ++(-2.0,0) |- (block6.west);
    % (block18) -| ++(2.5,0.0) node[pos=0.1,fill=white,inner sep=0]{No} |- (block20);

\draw[-latex] 
    (block20) edge node[pos=0.4,fill=white,inner sep=2pt]{No}(block21)
    % [->] (block7.west) -| ++(-2.0,0) |- (block6.west);
    (block20) -| ++(3.0,0.0) node[pos=0.1,fill=white,inner sep=0]{Yes} |- (block12);

\draw[-latex] 
    (block19) |- (block21);
    % [->] (block7.west) -| ++(-2.0,0) |- (block6.west);
    % (block20) -| ++(3.0,0.0) node[pos=0.1,fill=white,inner sep=0]{Yes} |- (block12);

\draw[-latex] 
    (block21) edge node[pos=0.4,fill=white,inner sep=2pt]{No}(block22)
    % [->] (block7.west) -| ++(-2.0,0) |- (block6.west);
    (block21) -| ++(4.1,0.0) node[pos=0.1,fill=white,inner sep=0]{Yes} |- (block2);

\draw[-latex] 
    (block12) edge node[pos=0.4,fill=white,inner sep=2pt]{No}(block13)
    % [->] (block7.west) -| ++(-2.0,0) |- (block6.west);
    (block12) -| ++(-3.0,0.0) node[pos=0.1,fill=white,inner sep=0]{Yes} |- (block21);

\end{tikzpicture}
\caption{Flowchart of the complete process.}
\label{flowchart}
\end{figure}
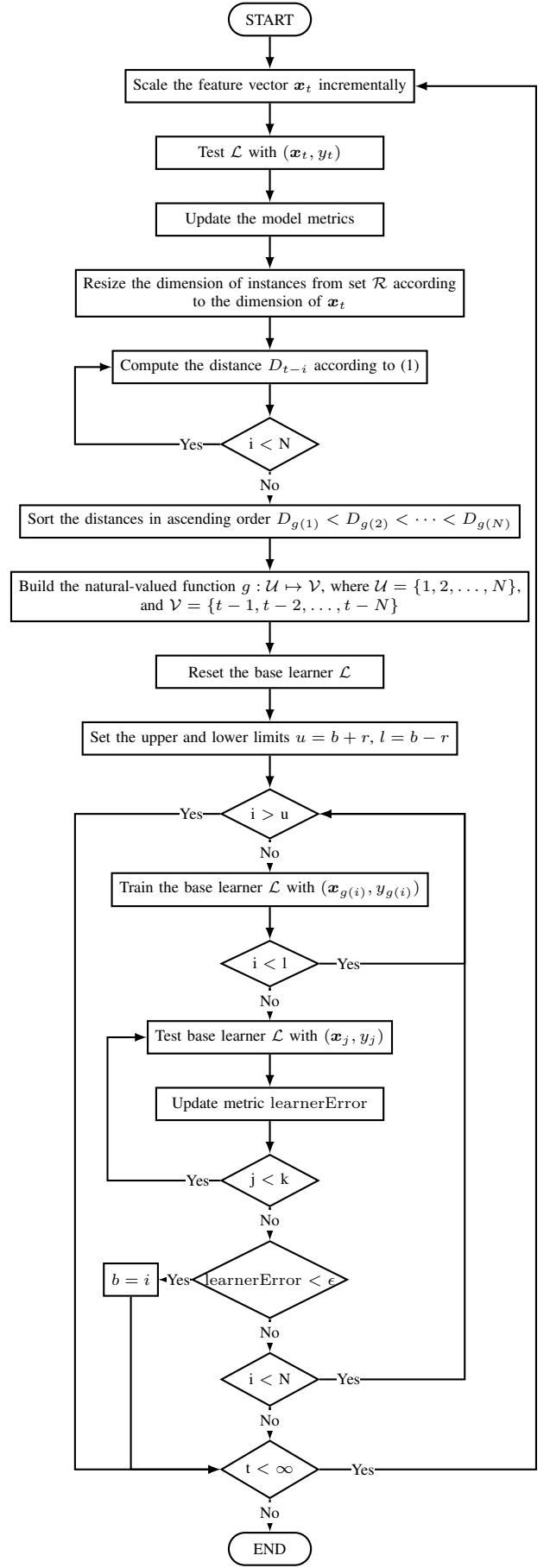

Then, SIS finds the optimal window $\mathcal{W}$ by solving:
\begin{mini}
  {\mathcal{W} \subset \mathcal{R}}{\lvert \mathcal{W} \rvert}{}{}
  \addConstraint{l \leq \lvert \mathcal{W} \rvert \leq u}
  \addConstraint{\frac{1}{k}\sum_{i=1}^k L_{\mathcal{W}}\big(\mathcal{L}(\bm{x}_{t-i}),y_{t-i} \big) \leq \epsilon }
  \addConstraint{L_{\mathcal{W}}= 
    \left\{\begin{matrix}
        1 & ;\:\mathcal{L}(\bm{x}_{k}) = y_{k}\\
        0 & ;\:\mathcal{L}(\bm{x}_{k}) \ne y_{k}\end{matrix}
    \right.}
  \addConstraint{\bm{x}_{g(j)} \in \mathcal{W},\quad \forall j\in\{1,\dots,|\mathcal{W}|\}}
  \label{sizesearchopt}
\end{mini}
Algorithms \ref{reordering} and \ref{setsearch} describe the procedures of SIS for solving problems \eqref{reorderingopt} and \eqref{sizesearchopt}, respectively. Fig. \ref{flowchart} shows the flowchart of the complete process.

\subsection{HAT and SIS Methods}
We propose the HAT+SIS learner, a variant of the Hoeffding Adaptive Tree for online power systems event detection. HAT is based on the Hoeffding Tree (HT) that establishes the Hoeffding bound to quantify the number of observations needed to compute some running statistics within a prescribed precision. Specifically, consider $n$ independent observations of a random variable of range $R$. The Hoeffding bound asserts that with high probability $(1-\delta)$ the estimated mean deviates from the true mean for no more than
\begin{align}
\epsilon = R\sqrt{\frac{\ln(1/\delta)}{2n}}.
\end{align}

HAT is composed of three main ingredients: a window to remember recent examples, the adaptive windowing (ADWIN) method as a distribution-change detector, and ADWIN as an estimator for some input data statistics. Once a change is detected, an alternate tree will be created and grow with the instances appearing right after the change. The alternate tree will replace the current tree if it is more accurate.

ADWIN serves as an estimator and change detector that keeps a variable-length window $\mathcal{W}$ of recent data such that the window has the maximal length statistically consistent with the null hypothesis that the average value inside the window has not changed. When two ``big enough" sub-windows of $\mathcal{W}$ have ``distinct enough" averages, it can be said with a high probability that a change in the data distribution has occurred and the older items in $\mathcal{W}$ should be dropped. The ``big and distinct enough" can be quantitatively defined by the Hoeffding bound \cite{Bifet2007}.
In \cite{Adhikari20182}, the authors introduce the HAT+DDM that is referred as HAT+ADWIN+DDM in \cite{Adhikari20182}. DDM is a concept drift detector based on statistical process control to detect changes in data streams. DDM considers two levels of the base learner's error rate: i) drift level when the error rate is very different from the past; and ii) warning level, when the error rate  has not reached the drift level. Once the error rate reaches the drift level, a new context is declared. Then, the base learner is trained with the instances between the warning and drift level exclusively; see details in \cite{Gama2004}.

%--------------------EXPERIMENTAL SETUP------------------------------
%------------------------------------------------------------------
\section{Experimental Setup}\label{exp}
We use Massive Online Analysis (MOA) \cite{MOA2010} and River \cite{River2020} to conduct the experiments. MOA is an open-source Java framework for streaming machine learning. River is an open-source python package dedicated to developing online/streaming machine learning algorithms. We run the experiments using a MacBook Pro 2019, 2.8 GHz Intel Core i7 processor, 16 GB 2133 MHz LPDDR3 RAM, and 1 TB hard disk drive. Based on initial tests for the cross-validation, we set the SIS hyperparameters as $N=200$, $k=1$, $r=10$ and $\epsilon = 0.1$.

\subsection{Testbed Architecture}
% The 3-bus power system with two generators shown in Fig. \ref{testbed} is modified from the IEEE 9-bus test case with three generators. The reduced system is small enough to comprehend its behavior in detail and apprehends the gist of the larger power system. 
The testbed architecture shown in Fig. \ref{testbed} comprises three main elements: physical power system, communication infrastructure, and monitoring and control components. The physical power system is simulated using the Real Time Digital Simulator (RTDS), which can emulate the behavior of power lines, buses, electrical machines, and load variations. The 3-bus power system with two generators shown in Fig. \ref{testbed} is modified from the IEEE 9-bus test case with three generators. The reduced system is small enough to comprehend its behavior in detail and apprehends the gist of the larger power system. The communication infrastructure comprises a physical network, industry-standard communication protocols, and control signals between control centers and substations. Also, networking monitoring devices such as SNORT and Syslog are used to track malicious network activity and log events or messages to the control panel. The monitoring and control components include hardware PMUs, PDCs, relays, a data processing engine, and industry-standard software. Each relay controls each breaker and sends information to the control panel through the communication network.  Please refer to \cite{Adhikari2014} for more details of the test bed architecture.

% The communication infrastructure in the test bed consists of  

% The hardware, software, and communication protocols used in the test bed are industry standard. 

% The RTDS provides an integration environment with software and hardware in the loop necessary to combine simulated power system with the communication and monitoring infrastructure. 

\subsection{Dataset}\label{datasets}
% The measurements are related to normal operation, electrical faults and cyberattacks collected from a power system, as shown in Fig. \ref{testbed}. There are three publicly available datasets randomly sampled from one initial dataset with 37 events. The datasets are categorized into two-class, three-class, and multi-class datasets. Each categorized dataset has 15 sets, where each of them contains around 5,000 instances. The binary dataset has two target classes: normal operation and attack events. The three-class dataset has three target classes: no events, natural events, and attack events. The target classes for the multi-class dataset are divided into 8 natural events, 1 case of no event, and 28 attack events. The datasets have 128 features: 116 PMU features of voltages, currents and impedances and 12 features of control panel logs, snort alerts and relay logs that are obtained from communication devices. In this work, we focus on the multi-class dataset, and the simulated events are listed as follows:
We test the three learners with the multi-class industrial control system (ICS) cyber attack dataset that includes measurements related to 37 events in an electric transmission system \cite{datasets}. The publicly available dataset consists of 15 sets with 5000 instances and 128 features each. The four PMUs measure 29 features each (voltages, currents, frequency, and impedances) for a total of 116 features. Also, 12 additional features correspond to information from SNORT, Syslog, and the control panel, totaling 128 features. The 37 simulated events are listed and distributed as follows:

\begin{figure}
    \centering
    \includegraphics[width=0.9\linewidth]{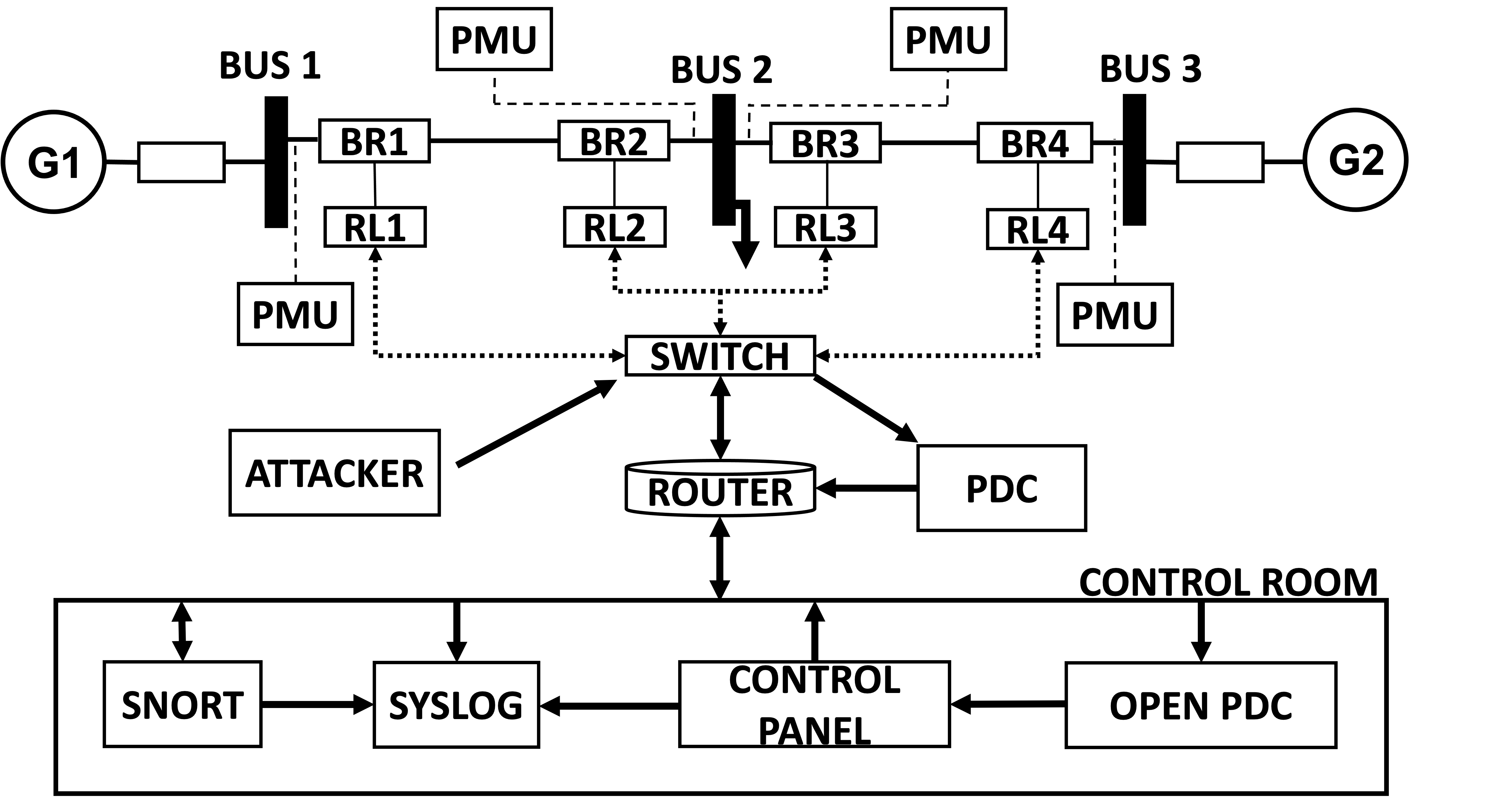}
    \caption{Cyber-physical power system testbed architecture. The circuit breakers and relays are shown as BR and RL respectively.}
    \label{testbed}
\end{figure}

\begin{itemize}
\item \textbf{Normal operation (1 event).} The system is operated under a stable condition with smooth load changes.
\item \textbf{Line maintenance (2 events).} The power lines are open via the protection relays.
\item \textbf{Short-circuit fault (6 events).} A power line short can occur in different locations across the line.
\item \textbf{Attack of remote tripping command injection (6 events).} An attacker opens a breaker by sending a command to a relay.
\item \textbf{Attack of relay setting change (16 events).} An attacker disables the secure relay configuration that forces the relay to not trip against real faults and valid commands.
\item \textbf{Attack of data Injection (6 events).} The attacker mimics a valid fault by changing the values of measurements such as voltages, currents, impedances, etc.
\end{itemize}

% The attack scenarios were simulated assuming that an intruder has access to the control room and can inject corrupted control commands to the network. The ICS datasets include a binary and ternary versions that are not

% Despite our proposed approach is the best performer using the binary (2 classes) and ternary (3 classes) versions of the ICS datasets

% We decided not to use the binary (2 classes) and ternary (3 classes) versions of the ICS datasets to assess the performance of the different classifiers due to space constraints.

\subsection{Performance Metrics}
The performance metrics used in this study are listed in blow:
\begin{enumerate}
    \item \textbf{Accuracy ($\%$):} The ratio between the number of correctly predicted instances and the total number of observed instances,
\begin{align}
\mathrm{Accuracy} = \frac{\mathrm{Number~of~correct ~predictions}}
{\mathrm{Number~of~observed~instances}}.
\end{align}  
    \item \textbf{Kappa ($\%$):} This statistic takes into account the possibility of predictions agreement by chance \cite{Zliobaite2014},
        \begin{align} 
        		\mathrm{Kappa} &= \frac{\rho_o - \rho_{\mathrm{ran}}}{1 - \rho_{\mathrm{ran}}},
        \end{align}
    where $\rho_o$ is the accuracy of the base learner under study, and $\rho_{\mathrm{ran}}$ is the accuracy of a random base learner. If $\mathrm{Kappa}$ is positive, the base learner is better than a random prediction. 
    \item \textbf{Time (s):} Processing time taken by the base learner.
    \item \textbf{Size (KB):} Size of the base learner in Kilobytes.
    \item \textbf{Cost (RAM-hour):} Amount of RAM memory (KB) deployed for one hour.
\end{enumerate}

\subsection{Learners}
We use the Hoeffding Tree (HT) \cite{Domingos2000} and variants of the Hoeffding Adaptive Tree (HAT) \cite{Bifet2009} as the base learners for the experiments. After an initial trial on the MOA classifiers for streaming machine learning, we select the HAT+DDM and HT+DDM learners because they show the better performance by using a portion of the multi-class dataset. We set the base learners with the hyperparameters suggested in the related literature. More detailed tuning of the base learner's hyperparameters is left for future work. HAT+SIS and HT+SIS exhibit the same performance for the multi-class dataset. Finally, we choose HAT+SIS because it has a better performance for a price forecasting dataset as shown in our preliminary experiments.

%--------------------CASE STUDIES------------------------------
%------------------------------------------------------------------
\section{Case Studies}\label{case_studies}
In this section, we assess the performance of the three learners under four scenarios with the multi-class dataset. The scenarios are simulated by imitating real fault disturbances and cyber-attacks. We explore those scenarios that affect the system's physics and architecture including e.g., loading variation, PMU disappearance, and measurement overlapping.

\begin{figure}[t]
    \centering
    \includegraphics[width=0.9\linewidth]{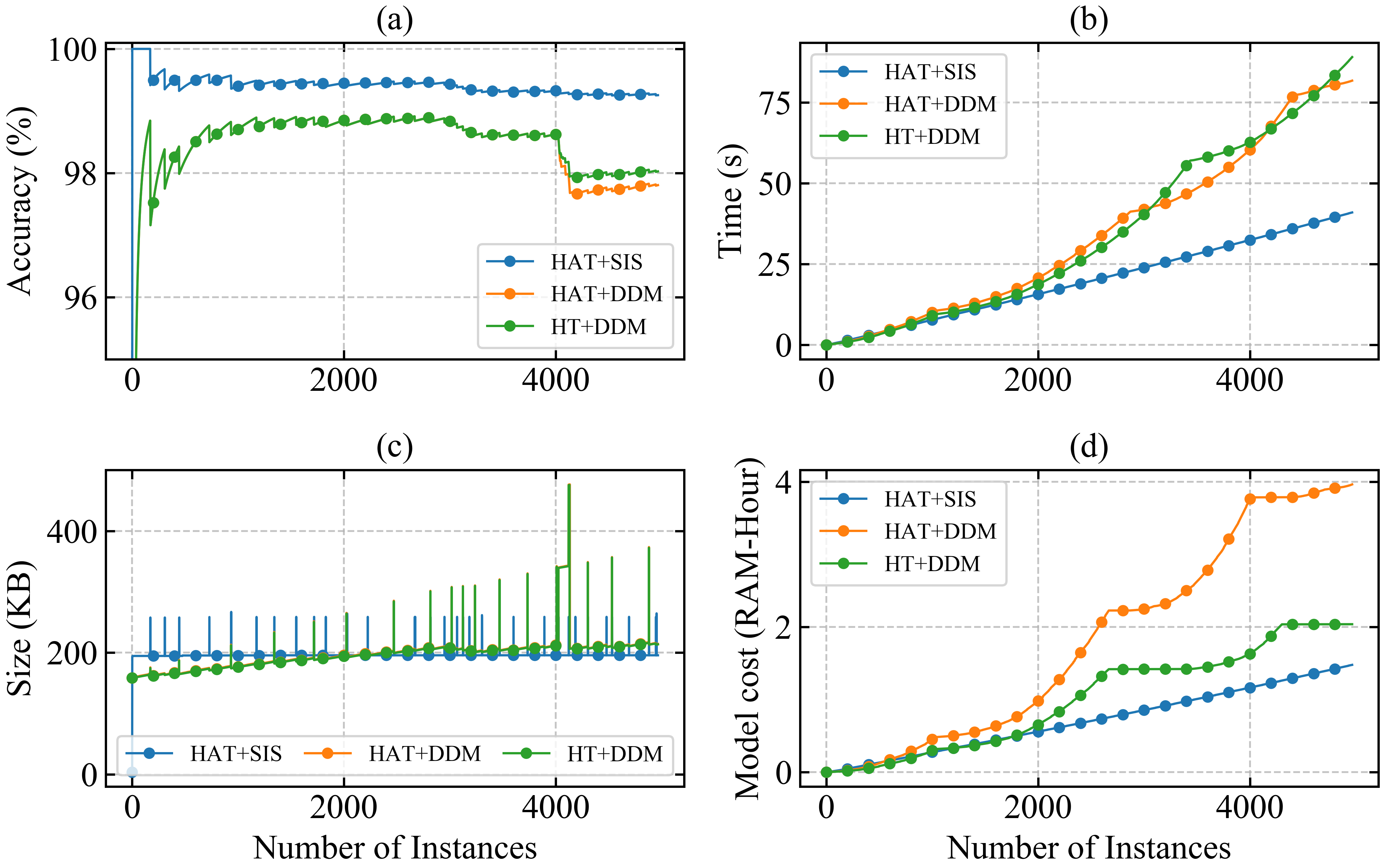}
    \caption{Performance comparison of the three learners under scenario I using the first set from the multi-class dataset. The evaluation considers the 37 events from the multi-class dataset in sequence. The comparison is shown for the following metrics: (a) Accuracy, (b) Time, (c) Size, and (d) Model cost.}
    \label{comparison-1}
\end{figure}

\subsection{Scenario I: Multiple Events}\label{scenario1}
This scenario evaluates the performance of the three learners with the 37 events from the multi-class dataset. The performance comparison is presented in Fig. \ref{comparison-1}. It can be seen that HAT+SIS is the best performer among all the learners, whereas HAT+DDM is the worst performer. As shown in Fig. \ref{comparison-1}(a), HAT+SIS achieves an accuracy of more than $99\%$ in the first 250 instances, while the accuracy of HAT+DDM and HT+DDM is less than $99\%$ during the same interval. Around the instance 4000, HAT+DDM and HT+DDM get an abrupt decrease in accuracy while SIS+HAT remains changed. The learners HAT+DDM and HT+DDM exhibit the same performance during the first 4,000 instances. Then, HT+DDM shows a slightly higher accuracy of recovery rate than HAT+DDM. Fig. \ref{comparison-1}(b) presents the time comparison of the learners. We notice that HAT+SIS maintains a linear running time as the stream progresses. Model sizes of the three learners are shown in Fig. \ref{comparison-1}(c). It can be seen that the three learners demand a moderate memory. HAT+DDM and HT+DDM use the same model sizes while HAT+SIS exhibits reduced peak memory demands. Fig. \ref{comparison-1}(d) shows the combined effect of the time and model size.

\begin{figure}[t]
    \centering
    \includegraphics[width=0.9\linewidth]{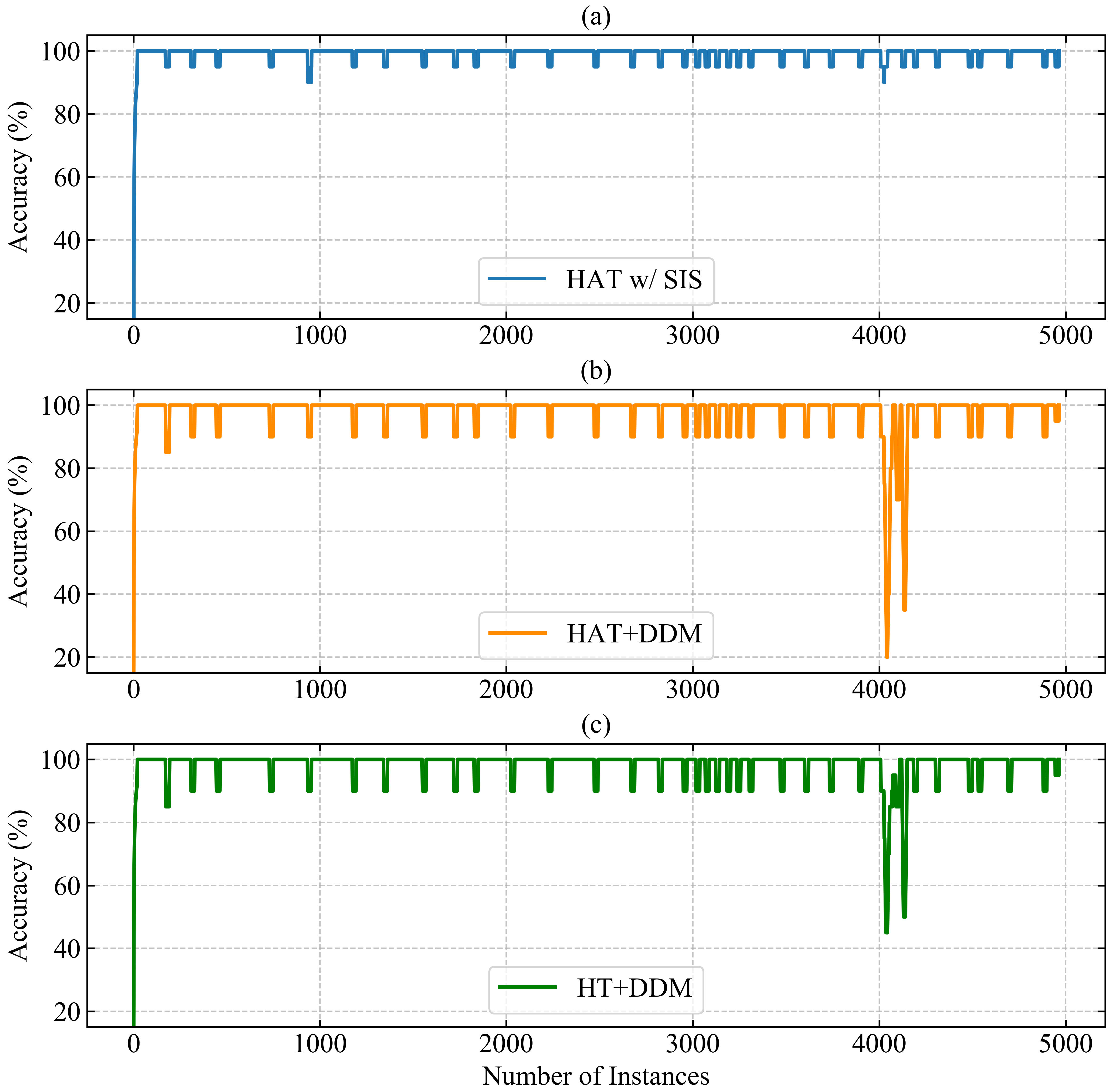}
    \caption{Performance evaluation of the base learners in dealing with stationary data and concept drift}
    \label{window-1}
\end{figure}

To evaluate the performance of the three learners both in static and evolving data, we use a window performance evaluator for classification with a window size of 20, as shown in Fig. \ref{window-1}. The three learners present the same accuracy during static data represented as horizontal lines reaching $100\%$ accuracy. Evolving data manifests as an abrupt or gradual decrease in accuracy displayed as downwards peaks. HAT+SIS exhibits the smallest peaks during evolving data, especially around instance 4000, where HAT+DDM and HT+DDM show a significant accuracy degradation.

The performance of the three learners among the fifteen sets from the multi-class dataset is shown in Table \ref{15datasets}. Considering the accuracy and Kappa statistic, HAT+SIS is the best performer among the fifteen sets, whereas HAT+DDM is the worst performer. Furthermore, HAT+SIS accounts for the least running time, model size, and cost. Table \ref{meanstd_datasets} presents the mean and standard deviation of the metrics across the fifteen sets. The results indicate that the performance of HAT+SIS remains invariant amidst the sets while exhibiting the most accurate and precise performance.

\begin{table*}[h]
\caption{Base learners performance for the 15 sets of the multi-class dataset.}
\resizebox{\linewidth}{!}{%
\begin{tabular}{c|c|c|c|c|c|c|c|c|c|c|c|c|c|c|c|c} 
\hline
Base        & Metric       & \multicolumn{15}{c}{Dataset} \\
\cline{3-17}
Learner     &               & 1 & 2 & 3 & 4 & 5 & 6 & 7 & 8 & 9 & 10 & 11 & 12 & 13 & 14 & 15 \\
\hline
%Learner & & $W = 24$ & $W = 8$ & $W = 4$ & $W = 2$ \\
\hline%.      Metrics.       				
			& Accuracy 		& 99.25     & 99.29 & 99.32 & 99.31 & 99.30 & 99.27 & 99.29 & 99.30 & 99.33 & 99.37 & 99.31 & 99.31 & 99.30 & 99.30 & 99.30	\\
 HAT		& Kappa			& 99.23     & 99.26 & 99.29 & 99.28 & 99.28 & 99.25 & 99.27 & 99.28 & 99.30 & 99.35 & 99.29 & 99.29 & 99.27 & 99.27 & 99.27	\\
 +   		& Time 	       	& 40.95     & 42.27 & 44.57 & 43.51 & 42.10 & 41.38 & 43.19 & 43.89 & 44.03 & 45.54 & 43.31 & 42.63 & 43.36 & 42.05 & 43.73	\\
 SIS  		& Size 	      	& 195.73    & 195.73 & 195.92 & 195.92 & 195.73 & 195.97 & 195.78 & 195.73 & 195.92 & 196.02 & 195.88 & 195.73 & 195.78 & 195.73 & 195.78  \\
         	& Cost  	& 1.48  & 1.53 & 1.61 & 1.57 & 1.52 & 1.50 & 1.56 & 1.59 & 1.59 & 1.64 & 1.56 & 1.54 & 1.57 & 1.52 & 1.58	\\
\hline%.      Metrics.       				
			& Accuracy 		& 97.81 & 98.42 & 96.66 & 98.02 & 98.57 & 98.51 & 98.05 & 98.61 & 97.98 & 97.36 & 98.59 & 98.58 & 98.60 & 98.55 & 97.16	\\
 HAT		& Kappa			& 97.73 & 98.37 & 96.53 & 97.94 & 98.52 & 98.46 & 97.98 & 98.55 & 97.89 & 97.25 & 98.54 & 98.53 & 98.54 & 98.50 & 97.04	\\
 +   		& Time 	       	& 81.71     & 76.35 & 75.52 & 129.42 & 78.41 & 99.14 & 86.84 & 84.07 & 211.72 & 210.14 & 169.43 & 88.73 & 84.13 & 164.01 & 72.02	\\
 DDM  		& Size 	      	& 214.34    & 217.94 & 228.26 & 221.93 & 220.38 & 214.81 & 223.03 & 224.87 & 226.26 & 231.94 & 223.19 & 222.14 & 224.54 & 219.10 & 224.17  \\
         	& Cost 	& 3.96  & 3.72 & 4.21 & 2.97 & 4.55 & 3.17 & 3.95 & 4.12 & 2.41 & 2.80 & 1.99 & 3.53 & 4.31 & 2.14 & 3.18	\\  
\hline%.      Metrics.       				
			& Accuracy 		& 98.03 & 98.42 & 97.67 & 98.21 & 98.57 & 98.51 & 98.17 & 98.61 & 97.98 & 97.86 & 98.59 & 98.58 & 98.60 & 98.55 & 97.82	\\
 HT 		& Kappa			& 97.96 & 98.36 & 97.58 & 98.14 & 98.52 & 98.46 & 98.10 & 98.55 & 97.89 & 97.78 & 98.54 & 98.53 & 98.54 & 98.50 & 97.73	\\
 +   		& Time 	       	& 89.00     & 69.63 & 66.30 & 84.05 & 65.36 & 92.21 & 78.98 & 62.95 & 161.98 & 162.21 & 194.52 & 228.09 & 70.49 & 142.79 & 81.59	\\
 DDM  		& Size 	      	& 213.68    & 217.12 & 227.43 & 220.94 & 219.38 & 213.65 & 222.03 & 223.88 & 225.10 & 230.94 & 222.37 & 221.14 & 223.54 & 217.94 & 223.18  \\
         	& Cost 	& 2.04  & 2.64 & 2.08 & 2.06 & 2.48 & 1.39 & 2.06 & 2.19 & 1.31 & 1.48 & 0.70 & 2.30 & 2.25 & 1.45 & 2.18	\\  
\hline
\end{tabular}%
}
\label{15datasets}
\end{table*}

\begin{table}[t]
\caption{The mean $\mu$, standard deviation $\sigma$, minimum and maximum values of the performance metrics tested on the 15 sets of the multi-class dataset.}
\begin{center}
\begin{tabular}{c|c|cccc} 
\hline
Learner        & Metrics       & \multicolumn{4}{c}{Results} \\
\cline{3-6}
     &               & $\mu$      & $\sigma$	& $\min$	& $\max$ \\
\hline
% %Learner & & $W = 24$ & $W = 8$ & $W = 4$ & $W = 2$ \\
\hline%.      Metrics.       				
		& Accuracy 	& 99.30	& 0.027	& 99.25	& 99.37	\\
 HAT		& Kappa		& 99.28	& 0.025	& 99.23	& 99.35	\\
 +   		& Time 	       	& 43.10     & 1.22	& 40.95	& 45.54	\\
 SIS  		& Size 	      	& 195.82   & 0.10	& 195.73	& 196.02	\\
         	& Cost 	& 1.56      	& 0.044	& 1.48	& 1.64	\\
\hline%.      Metrics.       				
         	& Accuracy	& 98.10	& 0.62	& 96.66	& 98.61	\\
  HAT      	& Kappa		& 98.02	& 0.64	& 96.53	& 98.55	\\
  +     	& Time 	       	& 114.11    	& 49.97	& 72.02 	& 211.72	\\
  DDM       	& Size 	      	& 222.46 	& 4.73	& 214.34 	& 231.94	\\
         	& Cost 	& 3.88     	& 1.01	& 1.99	& 4.55	\\
\hline%.      Metrics.       				
            	& Accuracy 	& 98.28   	& 0.33	& 97.67	& 98.61	\\
  HT        	& Kappa		& 98.21    	& 0.35	& 97.58	& 98.55	\\
  +         	& Time 	       	& 110.01   	& 53.52	& 62.95	& 228.09	\\
  DDM	& Size 	      	& 221.49   & 4.71	& 213.65	& 230.94	\\
            	& Cost 	& 1.81      & 0.494	& 0.70	& 2.64	\\
\hline
\vspace{-0.5cm}
\end{tabular}
\end{center}
\label{meanstd_datasets}
\end{table}

\subsection{Scenario II: Loading Variations}\label{scenario2}
\begin{figure}
  \begin{subfigure}{0.16\textwidth}
    \includegraphics[width=1.0\linewidth]{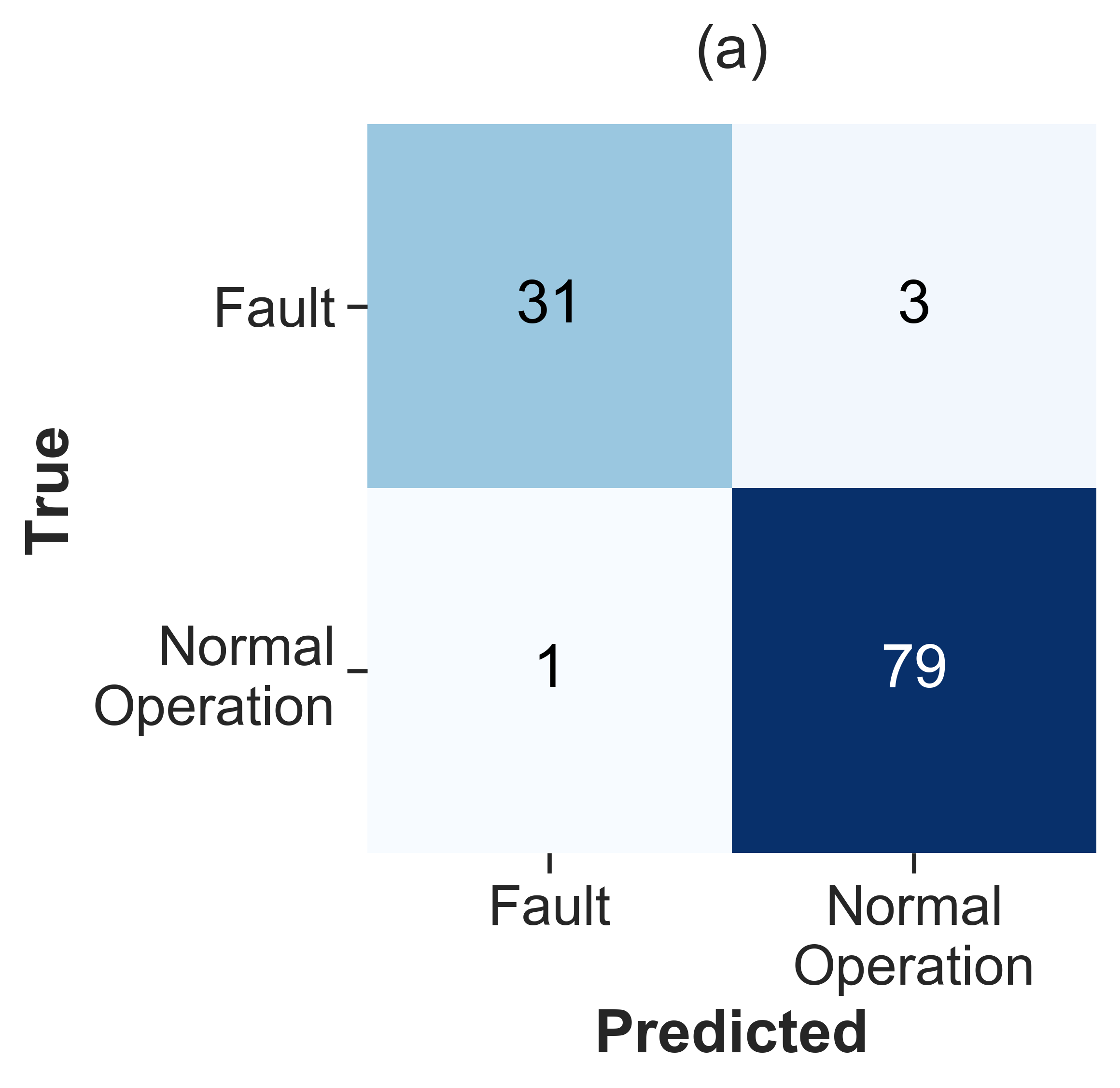}
    % \caption{}
    \label{fig:1a}
  \end{subfigure}%
  \hspace*{\fill}   % maximize separation between the subfigures
  \begin{subfigure}{0.16\textwidth}
    \includegraphics[width=1.0\linewidth]{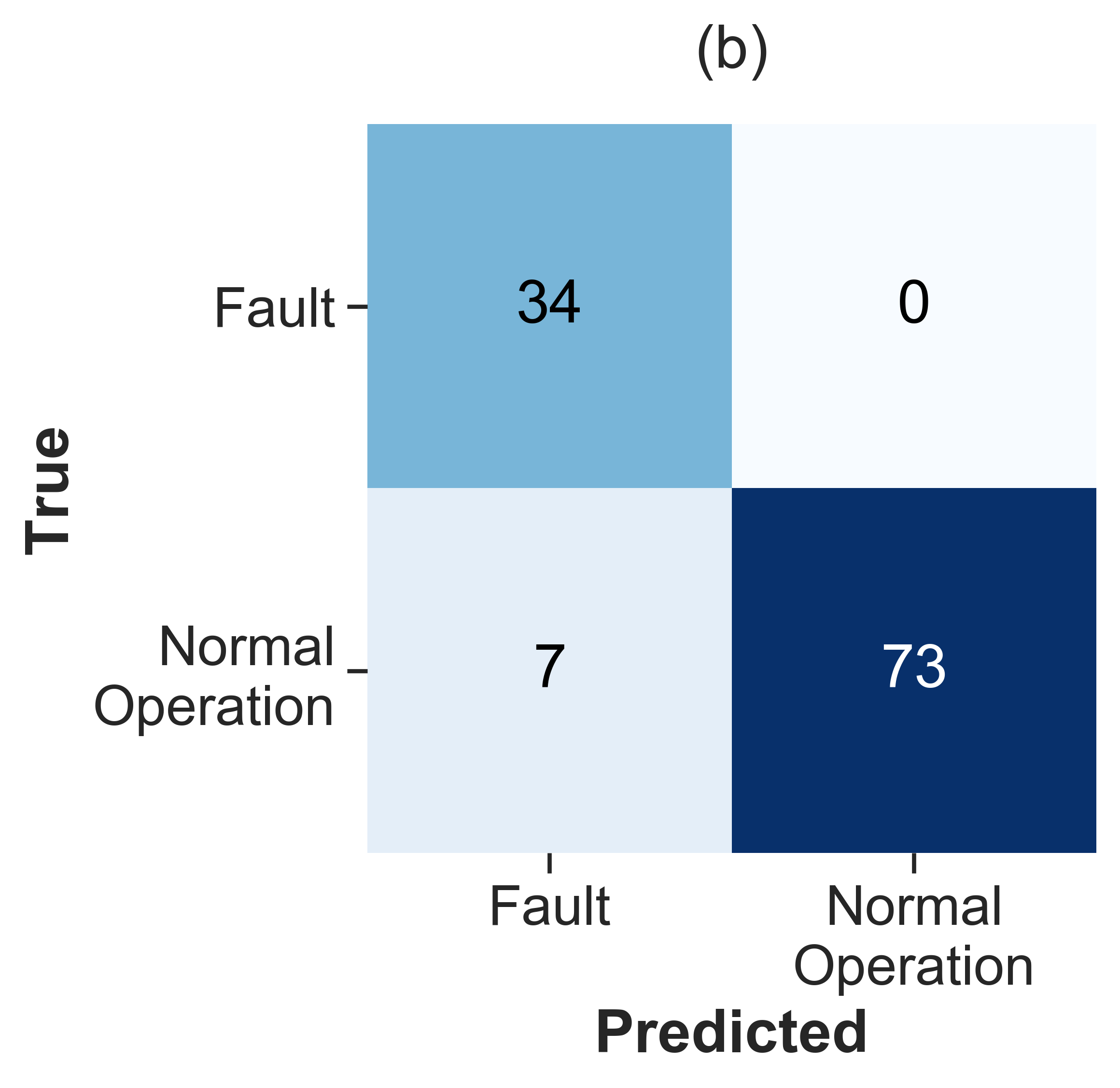}
    % \caption{}
    \label{fig:1b}
  \end{subfigure}%
  % \hspace*{\fill}   % maximizeseparation between the subfigures
  \hspace*{\fill}
  \begin{subfigure}{0.16\textwidth}
    \includegraphics[width=1.0\linewidth]{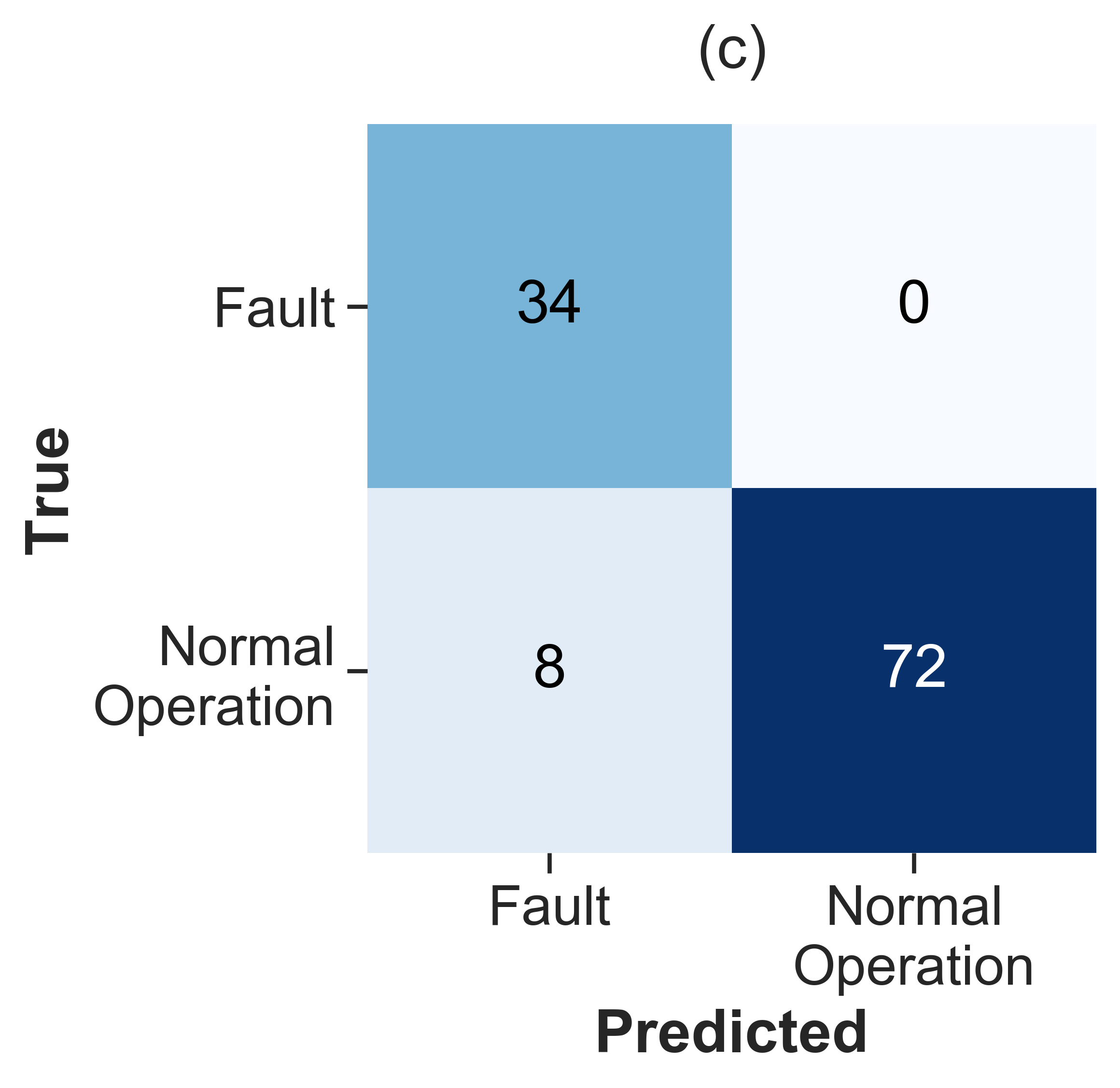}
    % \caption{} 
    \label{fig:1c}
  \end{subfigure}
  \hspace*{\fill}
\caption{Confusion matrices of the three learners in scenario II. (a) HAT+SIS, (b) HAT+DDM, (c) HT+DDM} 
\label{confusion}
\end{figure}

In this scenario, we test the three learners to classify a fault event under different loading conditions. We simulate a data stream that has three stages in this order: (i) fault from $10-19\%$ on line 2; (ii) the fault is cleared, returning the system to normal operation; and (iii) fault from $10-19\%$ on line 2. Stages (i) and (iii) are considered with distinct loads in the system. The confusion matrices are reported in Fig. \ref{confusion}. HAT+SIS accounts for the largest number of correct predictions while HT+DDM has the smallest. Though HAT+SIS classifies the instances from the normal operation event almost correctly, it performs the largest misclassifications for the fault event. HAT+DDM and HT+DDM have excellent performance for classifying the fault event. But they do not have an attractive performance for normal event classification.

\subsection{Scenario III: PMU Disappearance}\label{scenario3}
\begin{figure*}
\centering
  \begin{subfigure}{0.25\textwidth}
    \centering
    \includegraphics[width=1.1\linewidth]{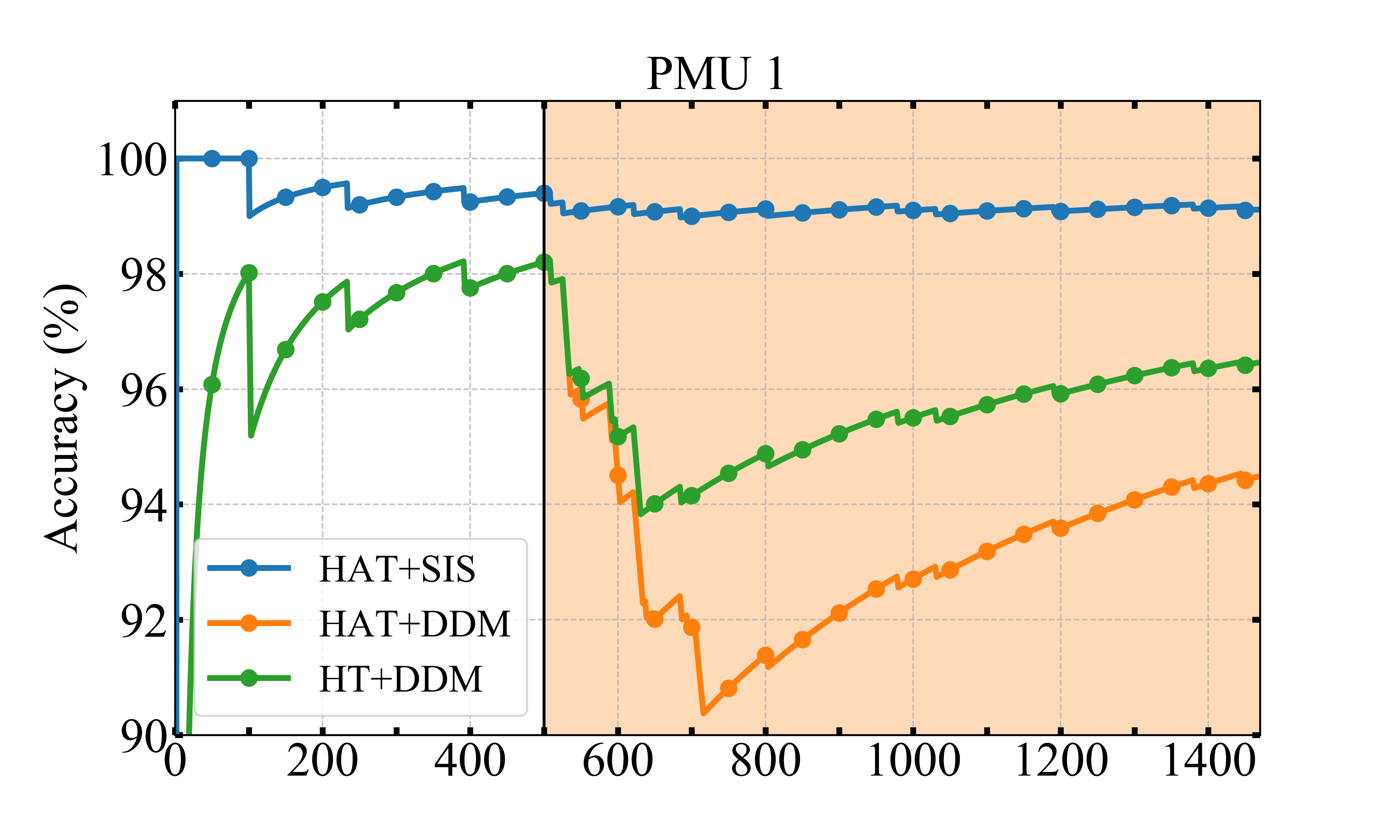}
    % \caption{}
  \end{subfigure}%
%   \hspace*{\fill}   % maximize separation between the subfigures
  \begin{subfigure}{0.25\textwidth}
    \centering
    \includegraphics[width=1.1\linewidth]{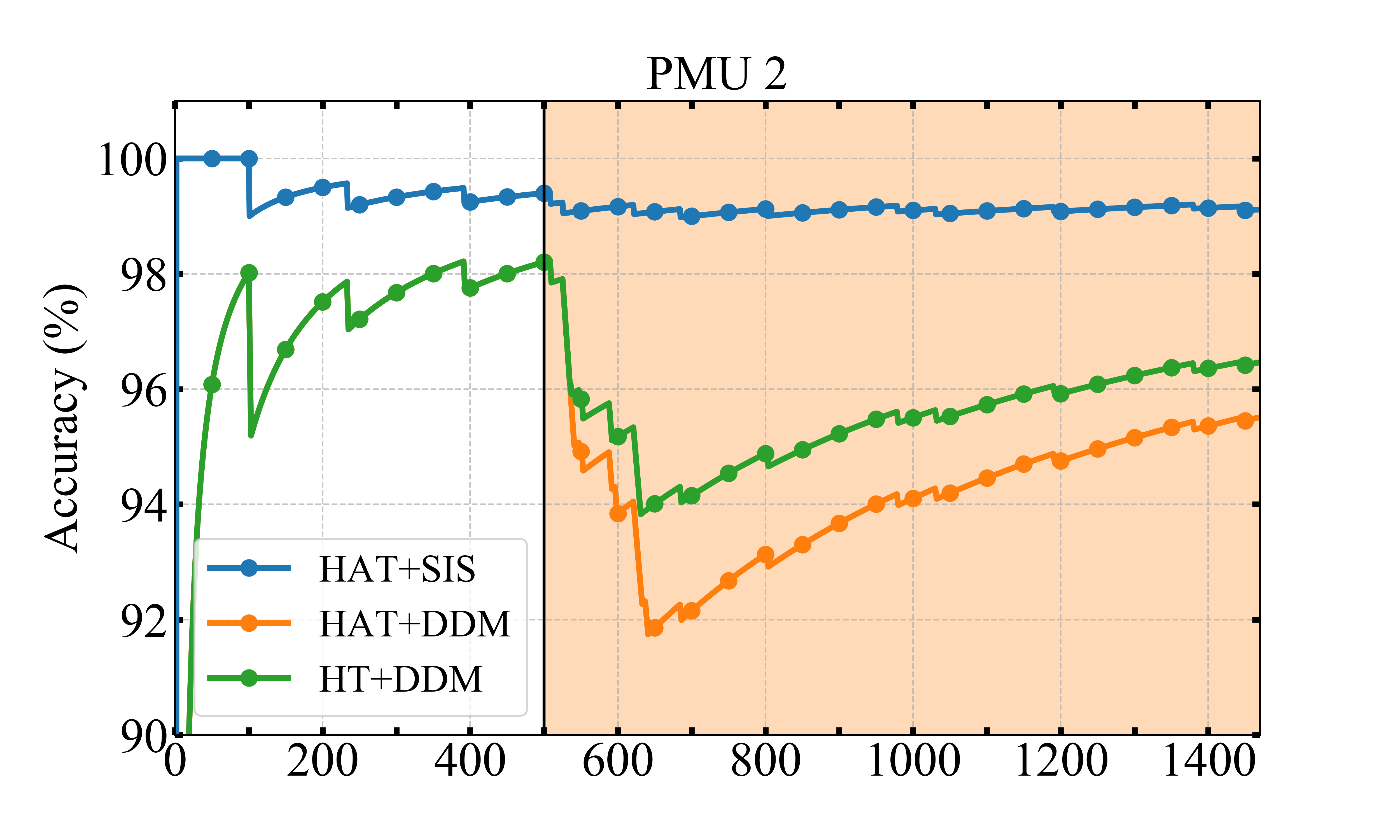}
    % \caption{}
  \end{subfigure}%
  \begin{subfigure}{0.25\textwidth}
    \centering
    \includegraphics[width=1.1\linewidth]{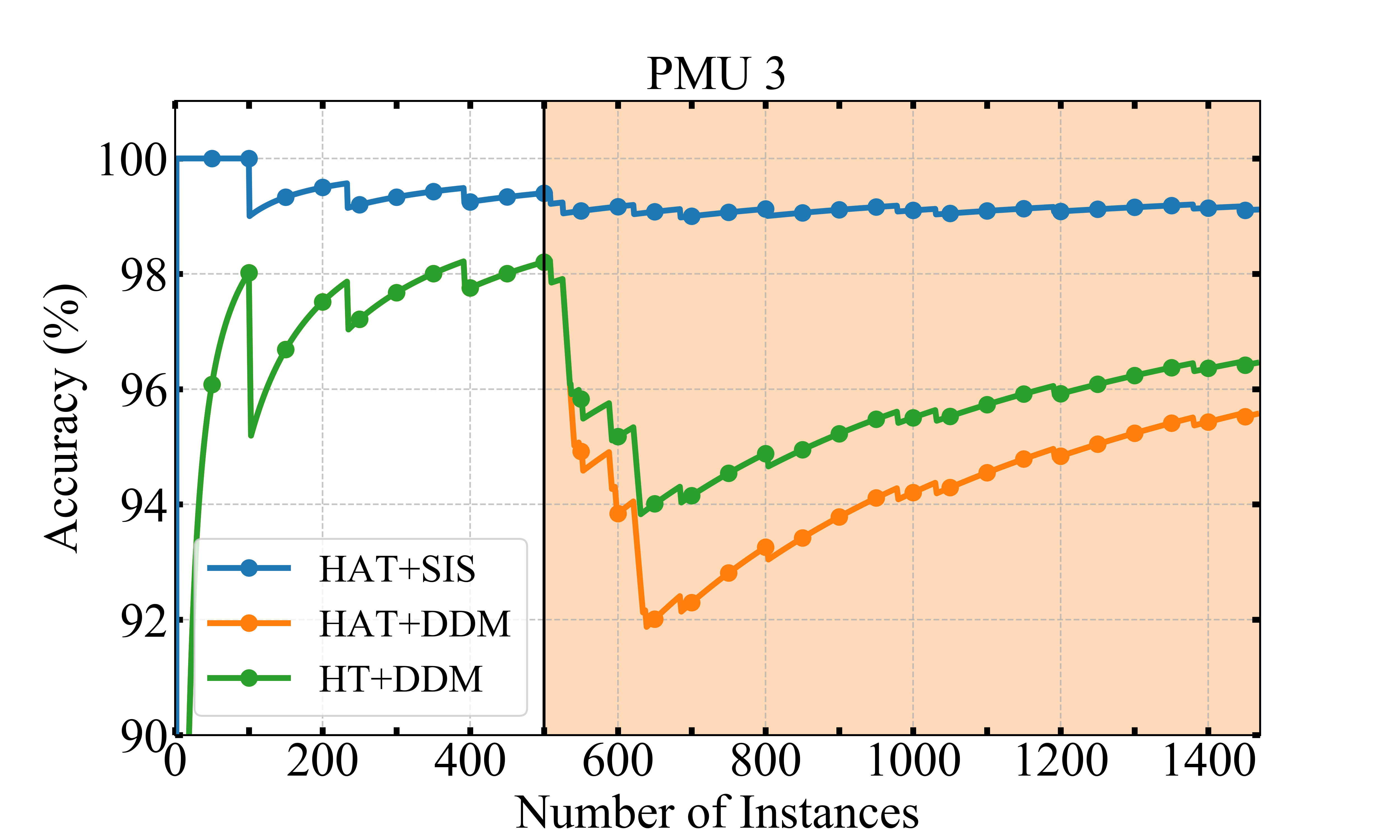}
    % \caption{} 
  \end{subfigure}%
%   \hspace*{\fill}   % maximizeseparation between the subfigures
  \begin{subfigure}{0.25\textwidth}
    \centering
    \includegraphics[width=1.1\linewidth]{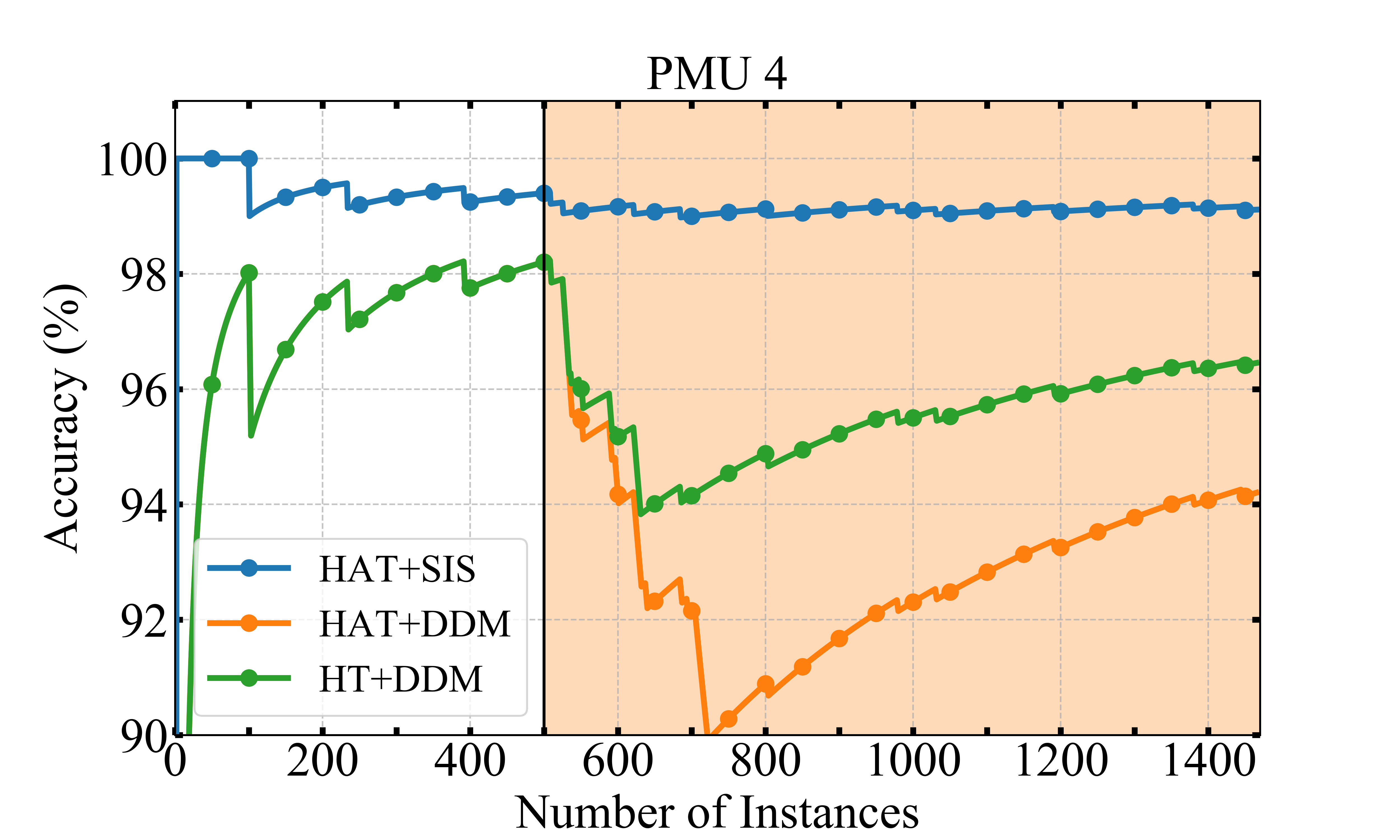}
    % \caption{} 
  \end{subfigure}%  
\caption{Accuracy of the three learners in scenario III. The vertical black bar indicates the instance at which one of the PMUs disappears in the data stream. The light orange shaded area corresponds to the portion of the data stream where the PMU is inactive, whereas the uncolored area corresponds to the section of the data stream where the PMU is active.} 
\label{PMU_Disappearance}
\end{figure*}

Consider a monitoring system consisting of four PMUs 1--4 whose measurements are modeled as features. After some time, one of the PMUs gets disconnected from the system. Such a situation can be modeled as a feature disappearance drift, which may occur due to a communication bottleneck, malfunctioning, or physical attack on the device. We simulate this scenario by using a data stream consisting of 1450 instances with line maintenance, remote tripping command, and fault events on both lines and different locations within the lines. Fig. \ref{PMU_Disappearance} shows the results for the three learners under scenario III with PMU's disappearance at instance 500. Clearly, HAT+SIS has the best performance.

\subsection{Scenario IV: Measurements Overlapping}\label{scenario4}
\begin{figure}
    \centering
    \includegraphics[width=0.9\linewidth]{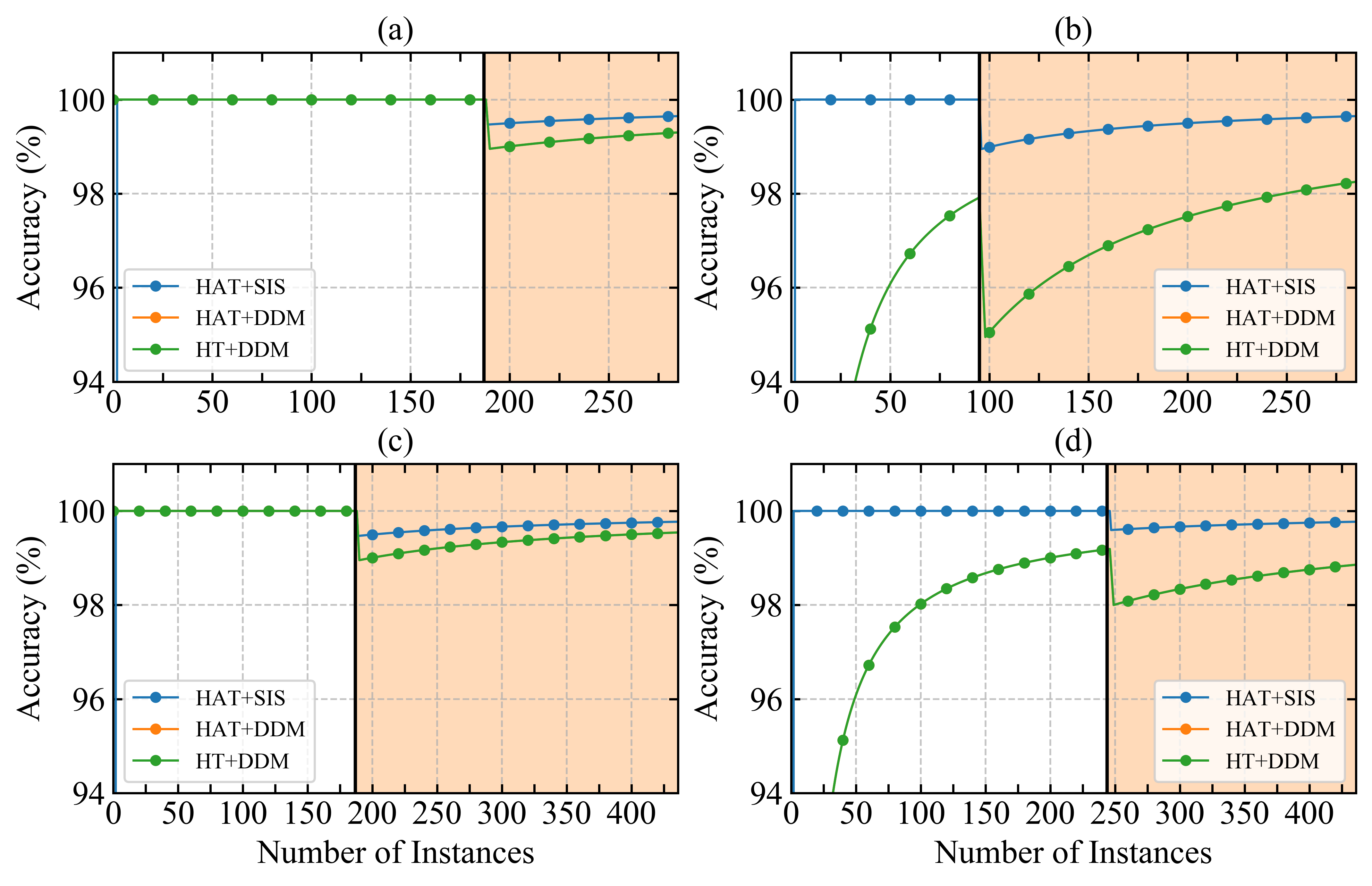}
    \caption{Accuracy of the three learners in scenario IV. The vertical black bar indicates the instance at which the abrupt change in the data stream occurs. The white shaded area corresponds to the portion of the stream under a fault disturbance distribution, whereas the light orange shaded area corresponds to a cyber-attack distribution. (a) and (b) for fault disturbance and remote tripping command events; (c) and (d) for fault disturbance and relay $\#2$ disabled. The performance of HAT+DDM and HT+DDM is the same in this scenario.}
    \label{Overlapping}
\end{figure}

Some cyber-attacks may exhibit similar class conditional measurements distribution $P(X|Y)$ as fault disturbances. In other words, cyber-attacks and faults may fall into the same region of the measurements (features) space. This makes it hard for learners to discriminate between similar events. In this scenario, we study one fault disturbance and two cyber-attacks: (i) fault from $80-90\%$ on line 1; (ii) a data injection attack that mimics a fault from $80-90\%$ on line 1 with remote tripping command; and (iii) a fault from $80-90\%$ on line 1 with relay $\#2$ disabled. First, we force the learner to process a data stream from disturbance (i) followed by an abrupt concept change in the data distribution corresponding to cyber-attacks (ii) or (iii). Then, we make the learner to learn oppositely, a cyber-attack (ii) or (iii) followed by the fault disturbance (i). Fig. \ref{Overlapping} shows the accuracy of the three learners under the scenario of measurements overlapping. It is evident that HAT+SIS is less vulnerable to the abrupt change in the data distribution. In Figs. \ref{Overlapping}(a) and \ref{Overlapping}(c), the three learners exhibit similar accuracy when they process instances from the fault distribution first. However, their accuracy performance are different if they start processing instances from the cyber-attack distribution, as seen in Figs. \ref{Overlapping}(b) and \ref{Overlapping}(d). The results indicate that HAT+SIS is the best performer in scenario IV because it can correctly classify instances from data injection and remote tripping attacks.

%--------------------DISCUSSION------------------------------
%------------------------------------------------------------------
\section{Discussion}
The case studies from section \ref{case_studies} show that HAT+SIS adapts to online monitoring environments and is robust against abrupt concept drifts. Both aspects suggest that HAT+SIS is suitable for real-time power systems events and intrusion detection classification systems. Besides the notable performance in accuracy and Kappa, one meritorious upshot of HAT+SIS is that its running time and model cost are considerably low compared with HAT+DDM and HT+DDM, as reported in Tables \ref{15datasets} and \ref{meanstd_datasets}. On the one hand, HAT+DDM and HT+DDM are very sensitive to concept drifts. Under certain conditions, their performance can be even worse than a random classifier, as shown in Fig. \ref{window-1}. On the other hand, HAT+SIS remains almost unaltered by such a concept drift, which avoids misleading insights about the current state of the system. Similarly, HAT+SIS shows a stable rate of correct predictions for classifying events with different loading conditions or with the sudden disappearance of a PMU as shown in Figs. \ref{confusion} and \ref{PMU_Disappearance}, respectively. HAT+SIS rapidly adapts its tree-based structure to discriminate similar events. Fig. \ref{Overlapping} shows the high accuracy recovery rate.

The 37 events allow us to judge the efficacy of any classifier, and a classifier that exhibits a remarkable performance using such events can be judged as a noteworthy classifier. Although the numerical results show that our proposed classifier exhibits a higher accuracy than other existing classifiers, as shown in Table \ref{comparisonAttack}, we know that the set of events is not exhaustive. We acknowledge that other events can be considered to strengthen our proposed approach. For instance, we can include denial of service attacks such as data flooding and mutation of MODBUS protocol or aurora attacks that refers to opening and closing a breaker near a generator in a rapid sequence.

\subsection{HAT+SIS Considerations}
Despite the merits of our proposed approach shown in Section \ref{case_studies}, a couple of conditions related to the HAT+SIS classifier need to be considered. The proposed method relies on reordering the instances based on their spatio-temporal distance to the target. Although instance reordering is an essential component of our proposed approach, it may degrade the original temporal distribution of the data stream, as shown in \cite{Zliobaite20112}. Data may acquire a different distribution over time, representing a different learning problem. Such a situation is a challenge because we aim to assess the performance of our classifier on a given real-time task while the data evolves unprecedently. In addition, SIS performs a window size search with a warm restart and stops the search only if the accuracy on a trial set is below a specified threshold. Such a strategy restricts the search around the previous best window size, leading to suboptimal window sizes. The search strategy imposes a trade-off between reducing the searching space and decreasing the method complexity not to violate time and memory restrictions.

\subsection{PMU Placement Strategy}
In this work, we use a testbed architecture of 3 buses with PMUs placed on all buses. If the test bed is a more extensive network, we may need to consider a specific PMU placement strategy for our proposed approach. Such a strategy will place PMUs in areas defined by clusters of buses that share the same dynamic behavior reducing the number of PMUs overall \cite{Cepeda2012}. The dataset considered for this work contains events from dynamic transients such as three-phase short circuits, line outages, load variations, and breaker tripping. For instance, placing a PMU in a region where a subset of the system’s buses exhibits similar behavior under a short circuit and line outages is sound. A drawback of this strategy is that it may be challenging for the classifier to identify where an event occurred among buses of the same cluster. Moreover, a tradeoff between the number of clusters and the cost of PMU deployment must be considered.

\subsection{Feasibility of the HAT+SIS Learner}
The learner must exhibit a fast response time for the event and intrusion detection task. As mentioned in section \ref{datasets}, the Attack dataset is built by using high-speed networking and PMU data. The PMUs make the data rate very high because they transmit tens or hundreds of synchrophasors per second through the networking architecture. This forces the learner to predict within a very short amount of time. In addition, the available working memory is not abundant, especially for PMUs or phasor data concentrators (PDC). From Table \ref{meanstd_datasets}, we notice that HAT+SIS processes approximately 5,000 instances from the Attack dataset in about 43 seconds. That is, the learner processes about 112 samples per second. We observe that the learner's model size is almost 196 KB. Such a memory demand can be handled by existing PMUs or PDCs in the market. Hence, HAT+SIS is a suitable classifier for PMU data based event and intrusion detection. 

\subsection{Comparison with Existing Works}
In Table \ref{comparisonAttack}, we compare the classification accuracy between the HAT+SIS learner and other algorithms based on the ICS dataset. Noticeably, we can observe that the HAT+SIS method outperforms the existing algorithms for classifying disturbances and cyber-attacks. The reported results for the HAT+SIS, HAT+DDM and HT+DDM learners are based on the average accuracy among the 15 sets from the multiclass dataset presented in Table \ref{meanstd_datasets}.

\begin{table}[H]
\caption{Accuracies of HAT+SIS and other adaptive classifiers on the Attack dataset reported in literature.}
\begin{center}
\begin{tabular}{l|c|l}
 \hline
Algorithm							& \# of classes	& Accuracy 	\\
\hline      
\hline
Common Path Mining \cite{Pan2015}		& 7			& 93.00	\\
NNGE+STEM \cite{Adhikari2018}		& 41			& 93.00	\\
HAT+DDM \cite{Adhikari20182}	& 41			& 92.00	\\
Weights Voting Algorithm \cite{Wang2019}& 37			& 92.40	\\
SSHAD \cite{Intriago2021}				& 37			& 96.84	\\
Tree-based GBFS \cite{Upadhyay2021}	& 37			& 92.46	\\
HAT+DDM							& 37			& 98.10	\\
HT+DDM							& 37			& 98.28	\\	
HAT+SIS         						& 37			& $\bm{99.30}$	\\
 \hline
 \vspace{-0.5cm}
 \end{tabular}
\end{center}
\label{comparisonAttack}
\end{table}

%--------------------CONCLUSION------------------------------
%------------------------------------------------------------------
\section{Conclusion}
This paper proposes a novel similarity-based streaming instance selection method for power system real-time events and intrusion classification. The similarity is based on our proposed spatio-temporal distance that uses an adaptive feature weight for PMU measurements with different scales. First, the SIS method reorders the most recent instances by similarity to the target instance. Then, using a sliding window, the method greedily searches for the most similar instances to the target instance. The search uses a warm restart to perform a local search around the previous best window size. We combine the SIS method with the HAT learner to create a robust steaming classifier that is strong against concept drift, such as fault disturbances and cyber-attacks. Moreover, we study the performance of the proposed classifier HAT+SIS along with other performers HAT+DDM and HT+DDM based on a dataset consisting of 37 power system events. We simulate various real situations, which may alter the physics of the system or the monitoring architecture. The experimental results show that our proposed learner can classify events in a power system with high accuracy of $99.30$ $\%$, low running time of $43.10$ $s$, model size of $195.82$ KB, and cost of $1.56$ RAM-hour.

% APPENDIX

\bibliographystyle{IEEEtran}
\bibliography{current/SIS_v5}

\end{document}